\documentclass[epj,onecolumn,nofootinbib,showpacs,showkeys]{revtex4-2}
\usepackage{epsfig}
\usepackage{psfrag}
\usepackage{amsfonts}
\usepackage{amssymb, bm}
\usepackage{amsmath, amsthm}
\usepackage{gensymb}
\usepackage{graphicx}
\usepackage{colordvi}
\usepackage{xcolor,colortbl}
\usepackage{dcolumn}
\usepackage{multirow}
\usepackage{tikz}
\usetikzlibrary{calc}
\usepackage{guit}
\usepackage[english]{babel}
\usepackage[T1]{fontenc}
\usepackage{lmodern}
\usepackage{epstopdf}
\usepackage{bm,natbib,url,textcase}
\usepackage{siunitx}

\PassOptionsToPackage{breaklinks}{hyperref}
\usepackage{hyperref}
\hypersetup{colorlinks,linkcolor={blue},citecolor={blue},urlcolor={blue}}

\usepackage{accents}
\usepackage{tabularx,siunitx}
\usepackage{makecell}
\usepackage{ctable}
\usepackage{wasysym}
\DeclareSIUnit\solarmass{M\ensuremath{_\odot}}
\usepackage{subfigure}

\usepackage{geometry}
\hypersetup{
	colorlinks=true,        
	linkcolor=blue,         
	citecolor=blue,         
	urlcolor=red           
}

\allowdisplaybreaks

\makeatletter

\renewcommand*{\p@subsection}{}

\renewcommand*{\p@subsubsection}{}
\makeatother

\begin{document}
\begin{flushleft} \centering Accepted for publication in EPJ C.
	\end{flushleft}

 \title{Relativistic periastron advance beyond Einstein theory: analytical solution with applications}

	\author{A. Tedesco$^{1}$\footnote{e-mail address: atedesco@unisa.it}, A. Capolupo$^{1,2}$\footnote{e-mail address: capolupo@sa.infn.it}, G. Lambiase$^{1,2}$\footnote{e-mail address: lambiase@sa.infn.it}}
	
	\affiliation{$^1$Dipartimento di Fisica ``E.R. Caianiello'',  Universit\`{a} degli Studi di Salerno, via G. Paolo II, I  - 84084 Fisciano, Italy}
 
	\affiliation{$^2$Istituto Nazionale di  Fisica Nucleare (INFN), Gruppo collegato di Salerno, via G. Paolo II, I - 84084 Fisciano, Italy}

\begin{abstract}
We find a new solution to calculate the orbital periastron advance of a test body subject to a central gravitational force field, for relativistic theories and models beyond Einstein. This analitycal formula has general validity that includes all the post-Newtonian (PN) contributions to the dynamics and is useful for high-precision gravitational tests. The solution is directly applicable to corrective potentials of various forms, without the need for numerical integration. Later, we apply it to the Scalar Tensor Fourth Order Gravity (STFOG) and NonCommutative Geometry, providing corrections to the Newtonian potential of Yukawa-like form $V(r)=\alpha \frac{e^{-\beta r}}{r}$, and we conduct the first analysis involving all the PN terms for these theories. The same work is performed with a Schwarzschild geometry perturbed by a Quintessence Field, leading to a power-law potential $V(r)=\alpha_q {r}^q$. Finally, by using astrometric data of the Solar System planetary precessions and those of S2 star around Sgr A*, we infer new theoretical constraints and improvements in the bounds for $\beta$. The resulting simulated orbits turn out to be compatible with General Relativity.
\end{abstract}

\date{May 21, 2024}
	
	

\maketitle

\section{Introduction}
Despite the unquestionable and numerous successes accomplished by General Relativity (GR) over the past century, observational studies have clearly shown that the dynamics of astrophysical objects at extragalactic scales are dominated by an invisible form of matter called dark matter. In particular, the effects of dark matter are manifest also at galactic scales, since rotation curves of galaxies show unexpected flat trends if Newtonian gravity is assumed with respect to the observed amount of baryonic emitting light matter. In addition to dark matter, the discovery that the universe is currently accelerating has led to the realisation that it is dominated by a form of energy of unknown origin and is supposed to be responsible for such a relevant phenomenon, called dark energy \cite{riess,ast,clo,spe,Carroll,sahini}. However, until now, no results have been obtained from the final experimental projects to detect particles that might form dark matter. However, if we give up the paradigmatic constraint that the gravitational anomalies observed at galactic and extragalactic scales are only caused by invisible matter composed of a new form of exotic particles, many other theoretical proposals can be taken into account. Another approach to understanding the nature of dark matter is represented by the Extended Theory of Gravity (ETG), whose paradigm follows Einstein's philosophy of curvature-based gravity field theory. The basic idea is that the Lagrangian density of the gravitational action (from which the field equation descends) is not simply the Hilbert-Einstein's one, i.e. a linear function of Ricci scalar, but a more general function of curvature invariants, possibly coupled on non-minimally coupled scalar field. For instance, including higher order invariants such as $\mathcal{L}=f(R)$ and $\mathcal{L}=f(R,\,R^2, \, R_{\mu\nu}R^{\mu\nu}, \,\Box R, \, \phi)$, that we can link to Einstein's gravity plus one or multiple coupled scalar fields by moving from the Jordan frame to the Einstein frame through a suitable conformal transformation \cite{Teyssandier:1983zz, Maeda:1988ab, Wands:1993uu, Schmidt:2001ac})  \cite{Berry:2011pb,Capozziello:2014mea, Lambiase:2015yia,Schellstede:2016ldu,NatureSci}. Other possibilities are given by the NonCommutative Geometry \cite{Lambiase:2013dai} which turns out to belong to the ETG class, and compactified extra dimension/Kaluza-Klein models \cite{ArkaniHamed:1998rs,ArkaniHamed:1998nn,Antoniadis:1998ig,Floratos:1999bv,Kehagias:1999my,Perivolaropoulos:2002pn}. Such a theoretical framework has aroused a growing interest in the scientific community, which lies in the fact that both dark matter and dark energy may be explained in a pure gravitational environment, and whose effects are interpreted to be provoked by the extra-curvature terms of spacetime. Importantly, one of the most relevant consequences is that the law of gravity has different strengths of attraction on different scales. The GR's gravitational pull is preserved in the Solar System, but in galaxies and clusters, it undergoes a variation of its strength due to growing contribution of extra-curvature terms. In other words, the gravitational pull is not scale-invariant, in agreement with the Mach principle.
The necessity of exploring the theoretical proposals inevitably leads to investigating the dynamics of celestial bodies in gravitating systems, and one of the most widespread and studied is the 2-body problem as a baseline for many astrophysical scenarios and tests of gravity. 

In this paper, we present the Scalar-Tensor Fourth Order Gravity (STFOG), which includes several sub-classes of ETG and NonCommutative Spectral Gravity (NCSG), and thus we consider the weak field limit providing new hypothetical forces, of which we aim to constrain the sizes. Correction of the Newtonian potential is in the form of a Yukawa-like function (5th force), that is, $V(r)=\alpha \frac{e^{-\beta r}}{r}$, where $\alpha$ is the parameter related to the strength of the potential and $\beta$ to the force range. We also take into consideration the Schwarzschild geometry deformed by a Quintessence Field, associated with the dark energy responsible for the present accelerated phase of the Universe and yielding a corrective power-law potential $V(r)=\alpha_q {r}^q$. Subsequently, we determine a general solution to calculate the relativistic periastron advance taking into account all kinds of perturbing potential terms. The final formula is found as a generalisation aiming to account independently of the form of the potentials, and it is based on the epicyclic perturbation method, on which Bertrand's Theorem was demonstrated \cite{Bertrand}, and is currently employed in the study of the physics of galaxies \cite{BinneyTremaine}; furthermore, such a formalism has already been successfully introduced for the GR perihelion advance (see R. Wald \cite{Wald} or \cite{Gravitation}), and has also been developed for some modified gravity models, first in ref. \cite{Harko} for Ho\v{r}ava-Lifshitz (HL) gravity, and then for the study of GR gravitational tests in the Solar System modified by the presence of a subdominant dark-matter halo \cite{Risi}. Putting together all these concepts, with the choice of the Binet (and Bertrand-like) approach \cite{Goldstein,Whittaker,Gravitation}, here we demonstrate how the epicyclic method can be performed to achieve a generalised \textit{analytical solution formula} for theories beyond Einstein that leads to a correction of the Newtonian potential, and how this leads to a straightforward calculation of the relativistic periastron advance. This is valid not only for the Solar System (or some binary star systems), where the orbit eccentricities of the main objects are small, but also for more general deviations from circularity (as occurs for the stars of the Sgr A* cluster). Such a solution is independent of the form of the corrective perturbing potential and is applicable to \textit{any} gravity field theory beyond GR, or a relativistic model within a certain theory, and it incorporates all post-Newtonian contributions with no need to involve numerical integration. First, we obtain the final analytic result and then deduce the expressions for the theories examined, through which new computations of the bounds can be performed, thus improving the results of our previous paper \cite{CapolupoLambiaseTedesco}. Finally, by taking advantage of the current astrometric data coming from the precession of planets, we analyse the effects of the post-Newtonian corrections to the periastron advance of planets in the Solar System and derive a lower bound on the adiabatic index of the equation of state. We proceed to infer constraints on the free parameter of the gravitational models. Non-Commutative Spectral Geometry (NCSG) is also studied, since it is a particular case of STFOG. Here, we show that our analytical results on the periaston advance of the planets, along with the S2 star around the Sagittarius A* Super Massive Black Hole at the centre of the galaxy, allow us to improve the bounds on the parameter $\beta$ by several orders of magnitude in this new work. Finally, such an analysis is studied to the case of power-law potential, referring in particular to the presence of a Quintessential Field around a Schwarzschild Black Hole, associated with the dark energy responsible for the present accelerated phase of the Universe. Before going on, we also mention that models of star orbits around the galactic centre in $f(R)$-gravity are investigated in \cite{DeLaurentis}, whereas for $f(R,\Box R)$-gravity one refers to \cite{Borka}.

In summary, the paper is organised as follows. In Section II we introduce Scalar-Tensor-Fourth-Order Gravity and NonCommutative Spectral Gravity as a particular class of ETG, then we show the weak field limit and the case relative to the Quintessence Field perturbing the Schwarzschild geometry. In Section III, we show the calculations by starting from the epyclic expansion and find an analytical solution for the relativistic periastron advance beyond Einstein theory, whose formula allows us to include all post-Newtonian contributions to the total precession. In Section IV, we perform a direct application and obtain the analytical results regarding STFOG, NonCommutative Spectral Gravity and Quintessence Field around a Schwarzschild Black Hole; this allows us to study the effects of the post-Newtonian corrections on the precession shift of planetary motions in the Solar System and of the S2 star motion around Sagittarius A*, and thus we derive new lower bounds on strengths and length of interaction of the Yukawa-like forces of the Extended Theories, as such as on the adiabatic index in the equation of state related to the power-law force due to the quintessential field. In Section V, we finally draw our conclusions with some remarks.

\section{Beyond Einstein Theory}
\subsection{Scalar Tensor Fourth Order Gravity}
As a general class representative of Extended Theories of Gravity (ETG), we consider the action for the Scalar-Tensor-Fourth-Order Gravity (STFOG) given by (see \cite{Felix})
\begin{eqnarray}\label{FOGaction}
\mathcal{S}\,=\,\int d^{4}x\sqrt{-g}\biggl[f(R,R_{\mu\nu}R^{\mu\nu},\phi)+\omega(\phi)\phi_{;\alpha}\phi^{;\alpha}+\mathcal{X}\mathcal{L}_m\biggr],
\end{eqnarray}
where $f$ is a generic function of the invariant $R$ (the Ricci scalar), $g$ is the determinant of metric tensor $g_{\mu\nu}$, the invariant $R_{\mu\nu}R^{\mu\nu}\,=\,Y$ ($R_{\mu\nu}$ is the Ricci tensor), $\phi$ is a scalar field and $\omega(\phi)$ is a generic function of it, $\mathcal{X}\,=\,8\pi G/c^4$. The Lagrangian density $\mathcal{L}_m$ is the minimally coupled Lagrangian density of ordinary matter.
The field equations are obtained by applying the variational principle to the action (\ref{FOGaction}) with respect to $g_{\mu\nu}$ and $\phi$. They read\footnote{We use, for the Ricci tensor, the convention
$R_{\mu\nu}={R^\sigma}_{\mu\sigma\nu}$, whilst for the Riemann
tensor we define ${R^\alpha}_{\beta\mu\nu}=\Gamma^\alpha_{\beta\nu,\mu} + \cdots$. The  Christoffel symbols are
$\Gamma^\mu_{\alpha\beta}=\frac{1}{2}g^{\mu\sigma}(g_{\alpha\sigma,\beta}+g_{\beta\sigma,\alpha}
-g_{\alpha\beta,\sigma})$, and we adopt the signature is $(+,-,-,-)$.}:
\begin{eqnarray}
\label{fieldequationFOG}
&&f_R R_{\mu\nu}-\frac{f+\omega(\phi)\phi_{;\alpha}\phi^{;\alpha}}{2}g_{\mu\nu}-f_{R;\mu\nu}+g_{\mu\nu}\Box
f_R+2f_Y{R_\mu}^\alpha
R_{\alpha\nu}+
\\\nonumber\\\nonumber
 &&-2[f_Y{R^\alpha}_{(\mu}]_{;\nu)\alpha}+\Box[f_YR_{\mu\nu}]+[f_YR_{\alpha\beta}]^{;\alpha\beta}g_{\mu\nu}+\omega(\phi)\phi_{;\mu}\phi_{;\nu}\,=\,
\mathcal{X}\,T_{\mu\nu}\,,\nonumber\\
\nonumber\\
\label{FE_SF}
&&2\omega(\phi)\Box\phi+\omega_\phi(\phi)\phi_{;\alpha}\phi^{;\alpha}-f_\phi\,=\,0 \,,
\end{eqnarray}
where:
\[f_R\,=\,\frac{\partial f}{\partial R}, \,\,\,\,\,\,f_Y\,=\,\frac{\partial f}{\partial Y}, \,\,\,\,\,\,\omega_\phi\,=\,\frac{d\omega}{d\phi}\,, \,\,\,\,  f_\phi\,=\,\frac{d f}{d\phi}\,, \]
and $T_{\mu\nu}\,=\,-\frac{1}{\sqrt{-g}}\frac{\delta(\sqrt{-g}\mathcal{L}_m)}{\delta
g^{\mu\nu}}$ is the the energy-momentum tensor of matter. We confine ourselves to the case in which the generic function $f$ can be expanded as follows (notice that the all other possible contributions in $f$ are negligible \cite{PRD1,FOG_CGL,FOGST})
\begin{eqnarray}
\label{LimitFramework2}
f(R,R_{\alpha\beta}R^{\alpha\beta},\phi)\,=\,&&f_R(0,0,\phi^{(0)})\,R+\frac{f_{RR}(0,0,\phi^{(0)})}{2}\,R^2+
\frac{f_{\phi\phi}(0,0,\phi^{(0)})}{2}(\phi-\phi^{(0)})^2\nonumber\\\\\nonumber&&+f_{R\phi}(0,0,\phi^{(0)})R\,\phi+
f_Y(0,0,\phi^{(0)})R_{\alpha\beta}R^{\alpha\beta}~.
\end{eqnarray}

\subsubsection{Weak field limit and solutions}
We are interested to solve the field equations for a non-rotating ball-like source of matter; thus the energy-momentum tensor reads
\begin{equation}
    T_{\mu\nu} = \rho (\mathbf{x}) c^2\, u_{\mu} u_{\nu} \,,\qquad T = \rho c^2\,,
\end{equation}
where $\rho(\mathbf{x}) c^2$ is the energy density of matter with $\rho(\mathbf{x})$ density of matter at rest, $c^2$ the square light speed, in the source's proper frame of reference $u^\mu$ fulfills the conditions $u^{\sigma}u_{\sigma} = 1$. In particular, for a ball-like source described as a perfect fluid without pressure, the components of the energy-momentum tensor are $T_{00}\,=\,\rho c^2$ and $T_{ij}\,=\,0$. The physical conditions of a static and weak gravitational field generated by a massive source (e.g. as it occurs in the Solar System), lead to study the weak-field limit of the theory. For Eqs. (\ref{fieldequationFOG}) and (\ref{FE_SF}), this means that we can search for solutions as expressions of the metric tensor $g_{\mu\nu}$ perturbing the Minkowski space-time $\eta_{\mu\nu}$ \cite{PRD1} as follows

\begin{eqnarray}\label{MeticCart}
&&g_{\mu\nu}\,\simeq\,
\begin{pmatrix}
1 + \overset 2g_{00}(x_0,\mathbf{x}) & 0 \\
0 & -\delta_{ij} + \overset 2g_{ij}(x_0,\mathbf{x}) \end{pmatrix}\,\,=\,\,
\begin{pmatrix}
1 + \dfrac{2}{c^2}\Phi(\mathbf{x}) & 0 \\
0 & -(1-\dfrac{2}{c^2}\Psi(\mathbf{x}))\delta_{ij}\end{pmatrix}\,,
\end{eqnarray}
and
\begin{eqnarray}
\nonumber
&&\phi\,\sim\,\phi^{(0)}+\phi^{(2)}+\dots\,=\,\phi^{(0)}+\varphi.
\end{eqnarray}

Through the overset number $2$ reported on the temporal and spatial components $\{\overset 2g_{00},\, \overset 2g_{ij}\}$ of the metric tensor, we recall that the related gravitational potentials $\{\Phi, \Psi\}$ and the scalar field $\varphi$ are of the order $c^{-2}$ in the post-Newtonian framework. Thus, by solving the resulting linearised version of field equations (\ref{fieldequationFOG}) and (\ref{FE_SF}) for a non-rotating source with radius ${\cal R}$ \cite{FOG_CGL,FOGST}, one obtains the following gravitational potentials and scalar field
%
\begin{eqnarray}\label{ST_FOG_FE_NL_sol_ball}
\Phi(\mathbf{x}) &=& -\frac{GM}{|\mathbf{x}|}\big[1 + g(\xi,\eta)\,e^{-m_+|\mathbf{x}|}+\Big[\frac{1}{3}-g(\xi,\eta)\Big]\,e^{-m_-|\mathbf{x}|}-
\frac{4}{3}\,e^{-m_Y|\mathbf{x}|}\big]\,, \label{PhiFOG}\\
\Psi(\mathbf{x}) &=& -\frac{GM}{|\mathbf{x}|}\big[1 - g(\xi,\eta)\,e^{-m_+|\mathbf{x}|} - \Big[\frac{1}{3}-g(\xi,\eta)\Big]\,e^{-m_-|\mathbf{x}|} - 
\frac{2}{3}\,e^{-m_Y|\mathbf{x}|}\big]\,, \label{PsiFOG} \\
\varphi(\textbf{x}) &=&\frac{GM}{|\textbf{x}|} \sqrt{\frac{\xi}{3}}\,\frac{2}{\omega_+-\omega_-} \biggl[\,e^{-m_+\,|\textbf{x}|}\, - \,e^{-m_-\,|\textbf{x}|}\biggr]\,, \label{varphiFOG}
\end{eqnarray}
where $f_R(0,0,\phi^{(0)})\,=\,1$, $\omega(\phi^{(0)})\,=\,1/2$, and
\begin{eqnarray}
g(\xi,\eta)\, &=& \,\frac{1-\eta^2+\xi+\sqrt{\eta^4+(\xi-1)^2-2\eta^2(\xi+1)}}{6\sqrt{\eta^4+(\xi-1)^2-2\eta^2(\xi+1)}}\,, \label{g-func} \\
\xi\,&=& \,3{f_{R\phi}(0,0,\phi^{(0)})}^2\,, \quad \eta\,=\,\frac{m_\phi}{m_R} \,, \label{m-functions}
\end{eqnarray}
while for the masses of Yukawa-like potentials in Eqs. (\ref{PhiFOG}) and (\ref{PsiFOG}), one has the relations
\begin{eqnarray}
m_\pm^2 \, = \, m_R^2\,w_\pm\,, \qquad w_\pm \, \, =  \, \, \frac{1-\xi+\eta^2\pm\sqrt{(1-\xi+\eta^2)^2-4\eta^2}}{2}\,,\label{mpm-w-functions}
\end{eqnarray}
with
\begin{eqnarray}
{m_R}^2\, \, \doteq \, \,-\frac{f_R(0,0,\phi^{(0)})}{3f_{RR}(0,0,\phi^{(0)})+2f_Y(0,0,\phi^{(0)})}\,,\qquad
        {m_Y}^2\, \, \doteq \, \,\frac{f_R(0,0,\phi^{(0)})}{f_Y(0,0,\phi^{(0)})}\,, \qquad {m_\phi}^2\,\doteq\,-\frac{f_{\phi\phi}(0,0,\phi^{(0)})}{2\omega(\phi^{(0)})}\,. \label{m-functions}
\end{eqnarray}

Since we are interested to the fields generated by \textit{ball-like source}, we remind that the Gauss theorem is satisfied only in General Relativity, where the exterior solution for a material point distribution coincide with the exterior solution for a generic spherically symmetric matter distribution. But in a Fourth Order Theory, this is no longer valid because of the Yukawa-like corrective terms in the potentials and a sphere cannot be reduced to a point. In this case, the equivalence no longer holds, and the type of distribution in the space is relevant. Therefore, in Fourth Order Theories, the Gauss theorem is not generally satisfied. In fact, if one considers a spherical mass with arbitrary density $ \rho(\mathbf{x}) $ and radius $\mathcal{R}$, the solutions relative to $\Phi$ and $\Psi$ show a geometric corrective factor that multiplies the Yukawa-like term depending on the form of the source \cite{PechlanerSexl,Stabile,CapolupoLambiaseTedesco}. For each term $\propto\,\frac{e^{-mr}}{r}$, this geometric factor is given by the function 

\begin{equation} \label{F-func}
F(m\,\mathcal{R})\,=\,3\frac{m\,\mathcal{R} \cosh m\,\mathcal{R} -\sinh m\,\mathcal{R}}{m^3\mathcal{R}^3}. 
\end{equation}
If we set $x = m\mathcal{R}$ in $F(m\mathcal{R})$, when $x \ll 1$, we have $\lim_{x\rightarrow 0} F(x) = 1$ so that a point-like source solution is recovered. In particular, the potentials in Eqs. (\ref{PhiFOG}) and (\ref{PsiFOG}) become
\begin{eqnarray}\label{Solutions-in-wfl-STFOG}
\Phi_{ball}(\mathbf{x})\, && =\,-\frac{GM}{|\mathbf{x}|}\biggl[1+g(\xi,\eta)\,F(m_+\mathcal{R})\,e^{-m_+|\mathbf{x}|} +
  \nonumber\\\nonumber\\
   && \qquad\qquad\qquad + [\frac{1}{3}-g(\xi,\eta)]\,F(m_-\mathcal{R})\,e^{-m_-|\mathbf{x}|} -
		\frac{4\,F(m_Y\mathcal{R})}{3}\,e^{-m_Y|\mathbf{x}|}\biggr], \label{Phi_ball}
		\nonumber\\\nonumber\\
		\Psi_{ball}(\mathbf{x})\, && =\,-\frac{GM}{ |\mathbf{x}|}\biggl[1-g(\xi,\eta)\,F(m_+\mathcal{R})\,e^{-m_+|\mathbf{x}|} - 
  \nonumber\\\nonumber\\
  && \qquad\qquad\qquad -[\frac{1}{3} -g(\xi,\eta)]\,F(m_-\mathcal{R})\,e^{-m_-|\mathbf{x}|}-
		\frac{2\,F(m_Y\mathcal{R})}{3}\,e^{-m_Y|\mathbf{x}|}\biggr]. \label{Psi_ball}
\end{eqnarray}
Some of the main ETG models studied in the literature, which can be summarised as subclasses of the more general STFOG, are reported in Table \ref{table} (see \cite{FOG_CGL} for other details). As we notice, generally the correction to the Newtonian potential is Yukawa-like with $V(r)= \alpha\frac{e^{-mr}}{r}$

\begin{center}
\begin{table*}[!ht]
\caption{\label{table} We report  different cases of Extended Theories of Gravity including a scalar field and higher-order curvature terms. The free parameters are given as effective masses with their asymptotic behaviour. Here, we assume  $f_R(0,\,0,\,\phi^{(0)})\,=\,1$,  $\omega(\phi^{(0)})\,=\,1/2$.}
{\small
\hfill{}
\begin{tabular}{|l|l|c|c|c|c|c|c|}
\hline
\multicolumn{1}{|c|}{\textbf{Case}}&\multicolumn{1}{|c|}{\textbf{ETG}}& \multicolumn{5}{ c |}{\textbf{Parameters}}  \\
\cline{3-7}
& & $m^2_R$ & $m^2_Y$ &$m^2_\phi$&$m^2_+$&$m^2_-$
 \\
\hline
A&\tiny{$f(R)$ }&\tiny{$-\frac{f_{R}(0)}{3f_{RR}(0)}$}&$\infty$& 0 & $m^2_R$ & $\infty$ 
\\
\hline
B&\tiny{$f(R,R_{\alpha\beta}R^{\alpha\beta})$}&\tiny{$-\frac{f(0)}{3f_{RR}(0)+2f_Y(0)}$}&\tiny{$\frac{f_{R}(0)}{f_Y(0)}$}& 0 & $m^2_R$ & $\infty$
\\
\hline
C&\tiny{$f(R,\phi)+\omega(\phi)\phi_{;\alpha}\phi^{;\alpha}$}&\tiny{$-\frac{f_{R}(0)}{3f_{RR}(0)}$}& $\infty$ &\tiny{$-\frac{f_{\phi\phi}(0)}{2\omega(\phi^{(0)})}$}&\tiny{$m^2_R w_{+}$}&\tiny{$m^2_R w_{-}$}
\\
\hline
D&\tiny{$f(R,R_{\alpha\beta}R^{\alpha\beta},\phi)+\omega(\phi)\phi_{;\alpha}\phi^{;\alpha}$} &\tiny{$-\frac{f(0)}{3f_{RR}(0)+2f_Y(0)}$}&\tiny{$\frac{f_{R}(0)}{f_Y(0)}$}&\tiny{$-\frac{f_{\phi\phi}(0)}{2\omega(\phi^{(0)})}$}&\tiny{$m^2_R w_{+}$}&\tiny{$m^2_R w_{-}$}
\\
\hline
\end{tabular}}
\hfill{}

\end{table*}
\end{center}

For our aims, as well as for many other astrophysical scenarios, it is more convenient (or simply required) to study models by resorting to spherical symmetry. For example, this is the case when the radial symmetry of the problem leads to central force fields, or the potentials are dependent on the mutual spatial distances between the positions of the bodies belonging to a given system or distribution of matter. It is readily possible to pass from space-time in \textit{isotropic coordinates} $x^{\alpha} = (x_0,x_1,x_2,x_3)$ 
\begin{equation}\label{isotropic-invariant}
    ds^2 = \biggl[ 1 + \dfrac{2}{c^2}\Phi(\mathbf{x})\biggr] c^2 dt^2 - \biggl[ 1 - \dfrac{2}{c^2}\Psi(\mathbf{x})\biggr]\delta_{ij} dx^i dx^j,
\end{equation}
to a spherically symmetric one $x^{\alpha} = (ct, r, \theta, \phi)$, by performing the transformation
\begin{equation}\label{transform-coord}
r^2 = [1-2\Psi(\mathbf{|x|})]|\mathbf{x}|^2,\qquad |\mathbf{x}| = x_i x^i\,,
\end{equation}
on the relativistic invariant (\ref{isotropic-invariant}) with potentials (\ref{Phi_ball}) and (\ref{Psi_ball}); working out the computations at first order with respect to the quantity $r_s/|\mathbf{x}|$ with $r_s = 2GM/c^2$ Schwarzaschild radius, we are able to find the STFOG space-time in spherical coordinates for a non-rotating ball

    \begin{eqnarray}\label{STFOG-spherical-metric}
     ds^2 && = \biggl[1-\frac{r_s}{r} \biggl(1 + g(\xi,\eta)\,F(m_+\mathcal{R})\,e^{-m_+ \, r} + [\frac{1}{3}-g(\xi,\eta)]\,F(m_-\mathcal{R})\, e^{-m_- \, r} - \frac{4 F(m_Y\mathcal{R})}{3}\,e^{-m_Y \, r} \biggl)\biggr] dt^2 - 
     \nonumber\\\nonumber\\
     &&  \qquad \quad- \biggl[1 + \frac{r_s}{r} \biggl(1 - g(\xi,\eta)(1 + m_+ \, r)\,F(m_+\mathcal{R})\,e^{-m_R \, r} - [\frac{1}{3} - g(\xi,\eta)] (1 + m_- \, r)\,F(m_-\mathcal{R})\, e^{-m_R \, r} - 
     \nonumber\\\nonumber\\
     && \qquad\qquad\qquad - \frac{2(1 + m_Y \, r)}{3}\,F(m_-\mathcal{R})\,e^{-m_Y r} \biggl)\biggr] dr^2 - r^2 d\theta^2 - r^2\sin^2\theta \, d\phi^2.
\end{eqnarray}

\subsection{NonCommutative Spectral Geometry}
NonCommutative Spectral Geometry (NCSG) is a special case of Scalar-Tensor-Fourth-Order Gravity, which is sparking growing interest in the scientific community as a theoretical candidate for the unification of all fundamental interactions, due to its intriguing properties \cite{ccm,ncg-book1,mairi2012} and offering a unique framework for studying several topics
\cite{Nelson:2008uy,Sakellariadou:2012jz,Chamseddine:2005zk,Chamseddine:2008zj,cchiggs,Chamseddine:2013rta,stabile}). Furthermore, satellite experiments allow us to identify precise constraints. At the scale of Grand Unification (fixed by the cutoff $\Lambda$), the Higgs field $\mathbf{H}$ is coupled to the gravitational sector of the action, and its variation with respect to $g_{\mu\nu}$ (see \cite{Nelson:2008uy,mairi2012}) yields the field equation 
\begin{equation}\label{2}
 G^{\mu\nu}+\frac{1}{\beta_{\small NCSG}^2}[2\nabla_\lambda \nabla\kappa
  C^{\mu\nu\lambda\kappa}+C^{\mu\lambda\nu\kappa}R_{\lambda\kappa}] = \mathcal{X}\,  T^{\mu\nu}\,,
\end{equation}
where $G^{\mu\nu}$ is the Einstein tensor, $\mathcal{X}\equiv 8\pi G/c^4$, $T^{\mu\nu}$ the energy-momentum tensor of matter and $\beta^{2}=\displaystyle{5\pi^2/(6 \mathcal{X} f_0)}$. A remarkable point is that neglecting the non-minimal coupling between the Higgs field and the Ricci curvature, the NCSG does not lead to corrections for homogeneous and isotropic cosmologies. This physical approximation enables us to analytically obtain a lower bound on $f_0$. By referring to the resolution of the linearised field equations, presented in \cite{LambiaseSakellariadouStabile,FOG_CGL} and achieved in harmonic coordinates, for the gravitational field potentials, one finds
\begin{eqnarray}\label{NCSG-solutions}
   \Phi(\mathbf{x}) &=&  - \frac{GM}{ |\mathbf{x}|}\left(1-\frac{4}{3}e^{-\beta
    |\mathbf{x}|}\right)\,, \label{NCSG-solution1} 
    \nonumber\\\nonumber\\ 
    \Psi(\mathbf{x})  &=&  - \frac{GM}{ |\mathbf{x}|}\left(1+\frac{5}{9}e^{-\beta
    |\mathbf{x}|}\right) \,. \label{NCSG-solution2}
 \end{eqnarray}
Performing once again the transformation (\ref{transform-coord}) on the metric tensor in isotropic coordinates (\ref{isotropic-invariant}) that originates from solutions (\ref{NCSG-solutions}), after computations, we obtain the following spherically symmetric space-time ($r_s= 2GM/c^2$)

\begin{eqnarray}\label{NCSG-spherical-metric}
    ds^2 = \biggl[1-\frac{r_s}{r} \biggl(1 - \frac{4}{3}\,e^{- \beta \, r} \biggl)\biggr] dt^2 - \biggl[1 + \frac{r_s}{r} \biggl(1 + \frac{5(1 + \beta \, r)}{9}\,e^{-\beta \, r} \biggl)\biggr] dr^2 - r^2 d\phi^2 - r^2\sin^2\phi \, d\theta^2.
\end{eqnarray}
\subsection{Quintessence Field: dark energy}
The Quintessence Field represents another interesting proposal to deal with problems of the Dark Universe. In particular, this is invoked to explain the speed-up of the present universe \cite{Jamil:2014rsa}. Quintessence may generate a negative pressure, and, being diffuse everywhere in the Universe, it is invoked to be the reason for the observed phase of positive cosmological acceleration \cite{Jamil:2014rsa} and it may also be present around a massive astrophysical object that warps the space-time around it \cite{Kiselev:2002dx}. Studies of quintessential black holes are also motivated by M-theory/superstring inspired models
\cite{Belhaj:2020oun}
(see \cite{Chen:2008ra,Toshmatov:2015npp,Ghosh:2015ovj,Jamil:2014rsa} for applications). The solution of Einstein's field equations for a static spherically symmetric quintessence surrounding a black hole in 4 dimensions is given by \cite{Kiselev:2002dx,Chen:2008ra}
  \begin{equation}\label{frmetric}
g_{\mu\nu}=\text{diag}\left[1-\frac{r_s}{r}-\frac{2\lambda}{r^{3\omega_Q+1}}, \, -  \dfrac{1}{\left(1-\frac{r_s}{r}-\frac{2\lambda}{r^{3\omega_Q+1}}\right)} , \, - r^2, \, - r^2 \sin^2 \phi \right]\,,
\end{equation}
where $r_s = 2GM/c^2$ is the Schwarzschild radius, $\omega_Q$ is the adiabtic index of the equation of state, $-1\leqslant\omega_Q\leqslant -\frac{1}{3}$, and $\lambda$ the quintessence parameter.
The cosmological constant ($\Lambda$CMD model) follows from (\ref{frmetric}) with $\omega_Q =-1$ and $\lambda=\Lambda c^2 /6$, leading to the components of the metric tensor
\begin{equation}
    g_{tt}=1-\frac{r_s}{r}-\frac{\Lambda c^2}{3} r^2 \,,\qquad g_{rr}=-\dfrac{1}{1-\frac{r_s}{r}-\frac{\Lambda c^2}{3}  r^2} \,.
\end{equation}
As we see, in this case, the corrective potential has the power-law expression
$
    V(r)\,=\,\alpha_q\,r^q\,.
$
\section{Solution formula for the periastron advance determination}
In this section, we elaborate a general analytical expression for the periastron advance in the 2-body problem, valid and applicable to any theory and model, e.g. ETG, Quintessence Field, but also Non-Local Gravity, GR plus Dark Matter, Anti-de Sitter solution, Reissner-Nordstrom solution, etc., independently of the fact that the solution of the field equation is exact or inferred in the Weak Field limit. The resolution is based on the mathematical idea of epicyclic expansion. Epicycles were first introduced by ellenistic mathematicians and astronomers to reproduce the retrograde observed motion of the planet on the celestial sphere \cite{BinneyTremaine}. The final dynamics results as the composition of oscillations around a point moving along the trajectory executed by the body. J. Bertrand \cite{Bertrand} used this approach to prove the Theorem stating that the only potentials yielding closed elliptical orbits are the Newtonian and elastic ones. Diverse applications to galactic dynamics and GR are provided in refs. \cite{BinneyTremaine,Wald,Gravitation}; for the first works in which the formalism is developed for modified gravity vacuum solutions (more specifically the Ho\v{r}ava gravity as a quantum gravity proposal) and GR plus a dark-matter halo, in order to study classical tests in the Solar System (where the planetary eccentricities are small), see refs. \cite{Harko,Risi}. 

We demonstrate here how it is possible, through the epicyclic-based formalism, to find an analytical solution that generalises the result to any form of potential and is also valid for systems where large deviations from circularity are present (e.g. stellar motions within the Sagittarius A* cluster). This will be applicable to ETG and other alternative/modified theories, represented by a final expression suitable for any other model\footnote{Besides Einstein's first derivation, commonly known techniques for the calculation of the precession shift in General Relativity were given by A. Eddington (we also mention T. Levi-Civita), as well as interesting known resolutions, were proposed by Whittaker, Robertson \cite{Robertson} (using Hamiltonian formalism), Chandraskhar (resorting to elliptical integrals) \cite{Chandrasekhar} and Weinberg \cite{Weinberg}. Commonly used techniques can be found in \cite{Carroll} and \cite{Gravitation}. Furthermore, Adkins \& MacDonell \cite{mcdell} established a method to treat the precession shift for a greater number of potentials and, as a result, a class of integrals with respect to the potential examined}. 

\subsection{Equation of orbit and epicyclic method}
From a physical point of view, the epicyclic method consists in the fact that an elliptic orbit can be produced exactly by a perturbation of the stable circular orbit. Since the stable circular trajectory of radius $r_0$ corresponds to the orbital solution relative to the point of minimum $r_0$ of the effective potential of a test particle which moves subject to a central force field as the one given a Schwarzschild post-Newtonian field (e.g. motion of a satellite around the Earth, or planet around the Sun), this technique involves that we need a Taylor series expansion around the minimum point of the gravitational effective potential. Especially, it allows us to incorporate all the post-Newtonian potentials descending from the entire theory, not only those related to General Relativity. Consistently with Bertrand's theorem and refs. \cite{Bertrand,Whittaker,Goldstein,Gravitation,Harko,Risi}, we decided to deal with the problem by resorting to the \textit{Binet equation of orbits}. Since we are considering the \textit{restricted} 2-body problem, i.e. we reduce to the model of a massive test particle moving on the geodesics of a (Schwarzschild-like) space-time warped by a central dominant non-rotating spherical mass. Such a method is based just on the assumption on the spherical symmetry of the model. Let us consider a generic spherically symmetric space-time 
\begin{equation}\label{spherical-metric}
     ds^2 = \biggl[1+\dfrac{2}{c^2}\Phi(r)\biggr] c^2dt^2 - \biggl[1-\dfrac{2}{c^2}\Psi(r)\biggr] dr^2 - r^2d\Omega \, ,
\end{equation}
where $d\Omega = d\phi^2 + \sin^2\phi \, d\theta^2$. Starting from the Lagrangian of the system
$2L \, = \,g_{\mu\nu}\dot{x}^{\mu}\dot{x}^{\nu},$
we impose the initial conditions $\dot{\phi}=0$ and $\phi = \pi/2$ on the metric (\ref{spherical-metric}), so that the motion is planar with respect to the coordinates $r$ and $\theta$.  For the Quintessence Field, the metric is given by Eq. (\ref{frmetric}), which automatically provides $\Phi (r) = \Psi (r)$. 
\begin{figure}[b!]
    \centering
    \includegraphics[width=0.77\linewidth]{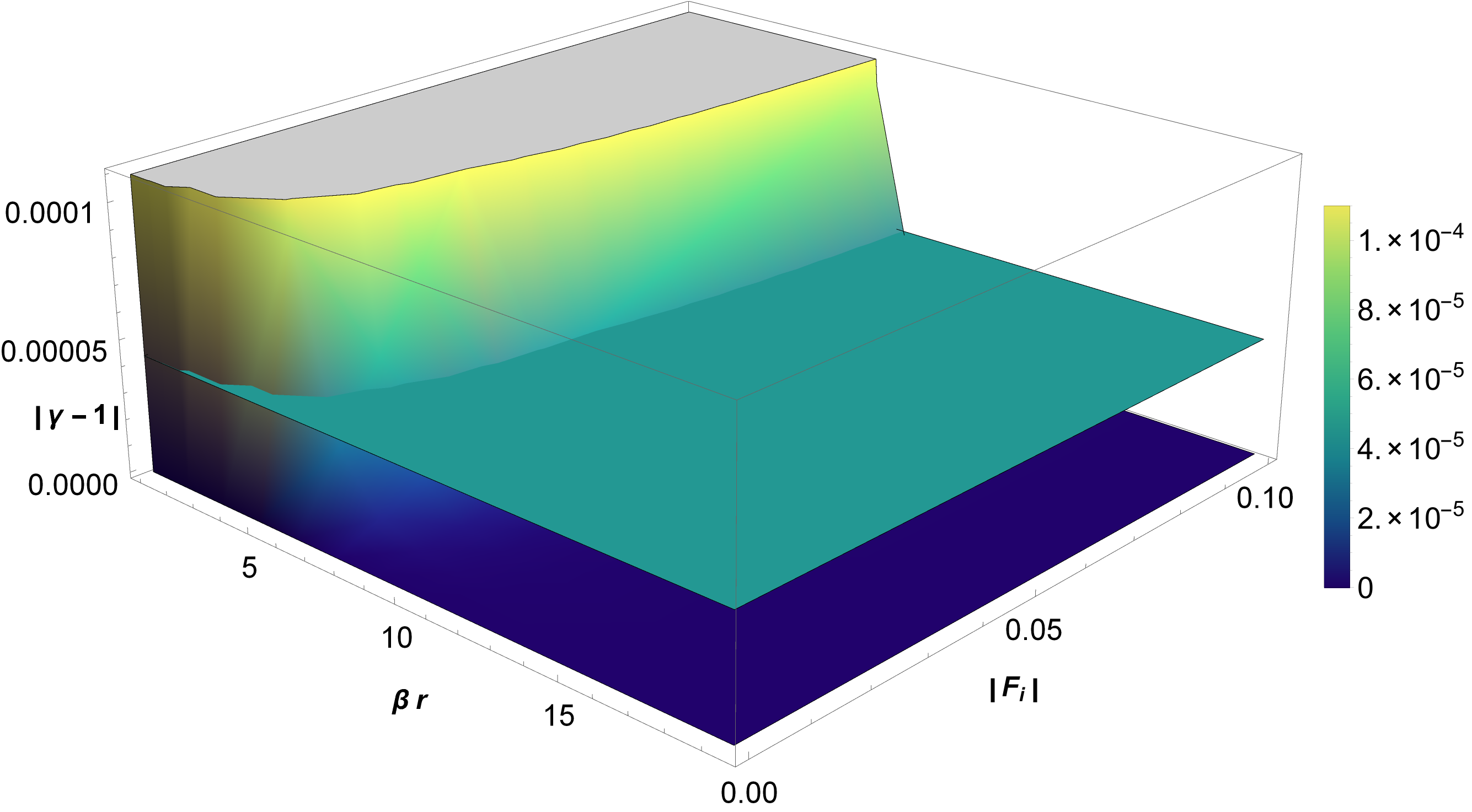}
    \caption{Plot of the function $|\gamma - 1|$ vs $\{ \beta r, |F_i| \}$, with $\gamma = \Phi / \Psi$ and $\{\Phi, \Psi \}$ given by Eqs. [\ref{Solutions-in-wfl-STFOG}] for STFOG. The part of the graphic below the plane $|\gamma - 1| = 4.4\times 10^{-5}$ corresponds to the domain satisfying the physical condition of small deviations from GR.}
    \label{fig:gamma-1}
\end{figure}
\begin{figure}[t!]
    \centering
    \includegraphics[width=0.55\linewidth]{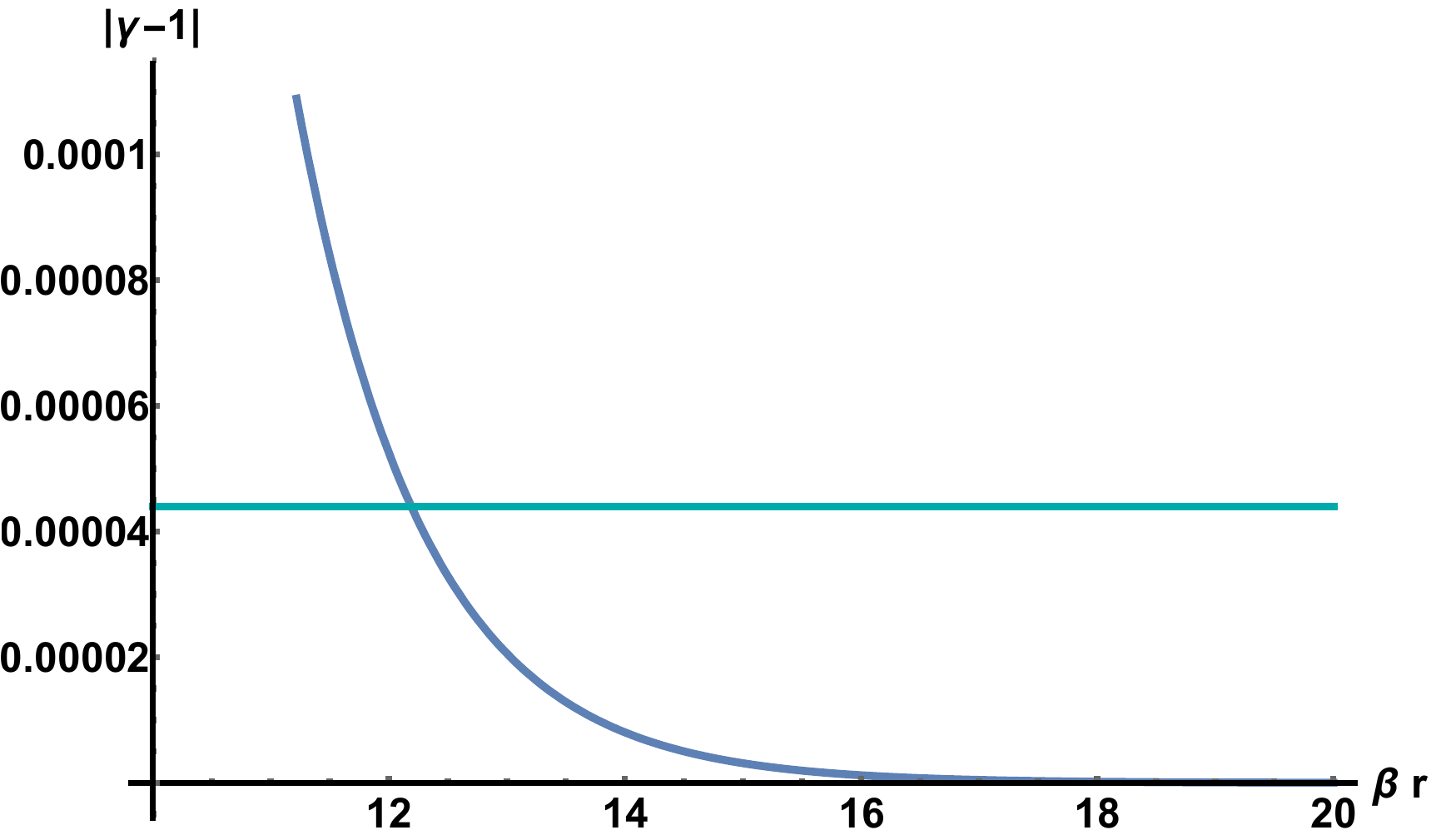}
    \caption{Plot of the function $|\gamma - 1|$ vs $\beta r$ (blue color), with $\gamma = \Phi / \Psi$ and $\{\Phi, \Psi \}$ given by Eqs. [\ref{NCSG-spherical-metric}] for NonCommutative Spectral Geometry (NCSG). The part of the graphic below the constant line $|\gamma - 1| = 4.4\times 10^{-5}$ (cyan color) corresponds to the interval satisfying the physical condition of small deviations from GR.}
    \label{fig:gamma-1}
\end{figure}
Concerning the ETG, we take into consideration the metrics given by Eqs. (\ref{STFOG-spherical-metric}) and (\ref{NCSG-spherical-metric}), arising from solutions (\ref{Solutions-in-wfl-STFOG}). Furthermore, we recall that in the framework of the Parametrized Post-Newtonian (PPN) formalism (see \cite{Eddington,Will,Gravitation}), a test of Einstein theory conducted in the Solar System, using radio links with the Cassini spacecraft \cite{Cassini}, obtained a tight constraint for the post-Newtonian parameter\footnote{It represents the quantitative contribution of the spatial part of the metric $g_{ij}$ to the space-time curvature.}
\begin{equation}\label{gamma}
    \gamma \, = \, \dfrac{\Psi}{\Phi}\, , \qquad  |\gamma_{\text{obs}} - 1| \, = \,  (2.1 \pm 2.3) \times 10^{-5}\,,
\end{equation}
in agreement with the General Relativity's value $\gamma=1$. According to Einstein theory, this value should also hold at the scales of Sgr A* stellar cluster, and therefore the departure from the GR's behaviour $\gamma \rightarrow 1$ (e.g. due to extra-curvature fields) must be very small. If we look at the potentials (\ref{Solutions-in-wfl-STFOG}), this occurs when the masses $m_+,\,m_-,\,m_Y$ of the gravitons associated with Yukawa-like interactions are large compared to the function $1/r$; while if these masses are small, we would find $\gamma \rightarrow 1/2$ at the examined scales. This value strictly violates the PPN parameter (\ref{gamma}), and such a situation would cause higher-order theories to be immediately ruled out. The same holds for the mass $\beta$ of the Yukawa-like interaction in NCSG (see Eq. (\ref{NCSG-solutions})). Being $\gamma=\Psi/\Phi$, if we look at the graphic for $|\gamma-1|$ (Fig. \ref{fig:gamma-1}) as a function of $\beta r$ and $|F_i|$, with $i=\pm,Y$ see Eq. (\ref{ST_FOG_FE_NL_sol_ball})), it is possible to identify the domain of physical parameters that satisfies the experimental constraint $|\gamma_{\text{obs}}-1|= 2.1\pm2.3\times 10^{-5}$. For example, if we consider $|F_i| \sim 10^{-1}$ and $\beta r \sim 10$, we see that the tiny percentage variation between the potentials $\{\Phi,\Psi\}$ is approximately $9\times10^{-6}$ \% from each other, meaning that $\Psi$ has a behaviour similar to $\Psi$.  Therefore, in such a domain, deviations from General Relativity turn out to be tiny, and the range of parameters dictated by Eq. (\ref{gamma}) is the one within the contribution of Yukawa-like potentials must exist (i.e. large $m_i$ as Yukawa-like interaction masses). Similarly, for NCSG (special case of ETG), the experimental constraint is satisfied for $\beta r \gtrsim 12.19$ and, if we consider the example value $\beta r \sim 14$, once again the variation between the potentials $\{\Phi,\Psi\}$ is very small, around $8.04\times10^{-6}$ \%. Now, we can start from the Lagrangian
\begin{equation}\label{lagrangian-expression}
 2 L  = [1+\dfrac{2}{c^2}\Phi(r)]c^2\dot{t}^2 - [1 - \dfrac{2}{c^2}\Psi(r)]\dot{r}^2 - r^2 \dot{\theta}^2.
\end{equation}
 where the dot indicates the derivative with respect to the proper time. The Euler-Lagrange equations
 \begin{equation}\label{E-L-equations}
\frac {d}{d\lambda} \frac{\partial L}{\partial \dot{x}^{\alpha}} - \frac{\partial L}{\partial x^{\alpha}}=0\,,
\end{equation}
with respect to coordinate time $t$ and the angle $\theta$, implies the constants of motion $E = [1 + (2/c^2) \, \Phi(r)] \, \dot{t}$,
and $h = r^2\dot{\theta}$,
which corresponds the conservation of energy (measured by a
static observer) and azimuthal angular momentum per unit mass of the test particle.
If we now insert these two relations into the first integral $g_{\mu\nu}\dot{x}^{\mu}\dot{x}^{\nu} \, = \,c^2$, one gets
\begin{equation}
    \dfrac{E^2}{\bigl[1+\dfrac{2}{c^2}\Phi(r)\bigr]}c^2 - [1 - \dfrac{2}{c^2}\Psi(r)]\dot{r}^2 -\dfrac{h^2}{r^2} \, = \, c^2\,,
\end{equation}
from which, after some computations, we have
\begin{equation}
    \dfrac{1}{2}\dot{r}^2 \biggl(1 +\dfrac{2}{c^2} \bigl[ \Phi(r) - \Psi(r)\bigr] - \dfrac{4}{c^4}\Phi(r)\Psi(r) \biggr) + \Phi(r) + \dfrac{h^2}{2 r^2}\biggl(1+\dfrac{2}{c^2}\Phi(r)\biggr) + \dfrac{1 - E^2}{2} c^2 \, = \, 0\, .
\end{equation}
By assuming $\Phi\sim\Psi$ due to the above physical considerations\footnote{Alternatively, since $\Psi = \gamma \, \Phi$, we also remark that $\Psi$ can equivalently be replaced with $\gamma \, \Phi$ in Eq. (\ref{lagrangian-expression}), where $\gamma \sim 1$ . In this way, we also notice that the negligible term $-\dot{r}^2 / c^2 \,(\gamma - 1) \lesssim 10^{-13}$ appears (with $|\gamma - 1| \lesssim 10^{-5}$), which is much smaller than the order $O(c^{-2})$, that is $\sim 10^{-5}$.} (whereas for the Quintessence field, the exact solution (\ref{frmetric}) automatically has $\gamma = 1$, i.e. $\Phi = \Psi$), and neglecting the higher order terms of the type $\sim \Phi \Psi \sim O(c^{-4})$ because irrelevant, the approximation leads the equation to be recast in the new form 
\begin{equation}
    \dfrac{1}{2}\dot{r}^2 + \Phi(r) + \dfrac{h^2}{2 r^2}\biggl(1+\dfrac{2}{c^2}\Phi(r)\biggr) + \dfrac{1 - E^2}{2} c^2 \, = \, 0.
\end{equation}
It is now possible to deduce the equation of motion in the suitable \textit{Binet form} by operating the substitution of the variable $u(\theta) \, \equiv \, 1/r$, from which follows $ \dot{r} = -h \, \dfrac{du(\theta)}{d\theta} $. Thus, 
\begin{equation}\label{Binet-equation}
    \biggl(\dfrac{du(\theta)}{d\theta}\biggr)^2 + u^2 \biggl[1 + \dfrac{2}{c^2}\Phi(u)\biggr] - \dfrac{2}{h^2}\Phi(u) + \dfrac{1 - E^2}{h^2} c^2 = 0 \, .
\end{equation}
Hereafter, we appropriately divide the potential $\Phi$ into the sum of two separate contributions as $\Phi = \Phi_N \, + \, \Phi_p$,
that is, into the usual Newtonian potential and the perturbing Yukawa-like potential. Then, differentiating with respect to $\theta$ this equation, we finally obtain the \textit{relativistic Binet equation of orbit}
\begin{equation}\label{Binet-equation}
    \dfrac{d^2u}{d\theta^2} + u = \dfrac{GM}{h^2} + \dfrac{3GM}{c^2} u^2 - \dfrac{1}{h^2} \, \Phi'_p (u) - 2 \dfrac {u}{c^2} \, \Phi_{p}(u) - \dfrac{u^2}{c^2} \, \Phi'_p (u)\,,
\end{equation}
 where the prime denotes the derivative with respect to $u$. Since the second member can be identified with the function
 \begin{equation}\label{J-function}
    J(u) \, =  \dfrac{GM}{h^2} + \dfrac{3GM}{c^2} u^2 - \dfrac{1}{h^2} \, \Phi'_p (u) - 2 \dfrac {u}{c^2} \, \Phi_{p}(u) - \dfrac{u^2}{c^2} \, \Phi'_p (u)\,,
\end{equation}
 the differential equation equation reads 
\begin{equation}
\dfrac{d^2u}{d\theta^2} + u = J(u)\,, 
\end{equation}
where we recognize $J(u) \, = \, - h^{-2} \, V'_e(u)$
as the function associated to the derivative with respect to $u$ of the effective gravitational potential (third, fourth, fifth terms at first member of Eq. [\ref{Binet-equation}]) multiplied by $-h^{-2}$, expressed by the second member of Eq. (\ref{Binet-equation}). We rapidly notice that the first term at the second member leads to the classical elliptic orbit of Newtonian gravity, while the second term is the post-Newtonian contribution of General Relativity to the central force leading to the Rosette orbit arising from the rotation of the apsidal line. The fourth, fifth, and sixth terms represent the post-Newtonian Yukawa contributions of the ETG to the dynamics. Now we apply the epicyclic perturbation: since the circular motion of radius $u_0 = 1/r_0$ occurs at the point of minimum of the effective potential, namely, the potential is such that the motion is stable and the solution $u$ results bounded also after a small variation from $u_0$, in order to describe the elliptic orbit, we add a slight perturbation so that 
\begin{equation}\label{epi-perturbation}
u = u_0 + x\,,
\end{equation}
with $u_0  =  GM/h^2  =  [a(1-\epsilon^2)]^{-1}$ obtained from the equation $u_0  = GM/h^2 \, + \, (3GM/c^2)\, u^2_0$.

\subsection{Solution for small deviations from circularity}
Inserting the relationship (\ref{epi-perturbation}) into the differential equation and expanding $J(u)$ in the Taylor series around $u_0 = GM/h^2$ as

\begin{equation}\label{taylorseries1}
J(u) \simeq J(u_0) + J'(u_0) x\,,
\end{equation}
where $J'(u_0)$ is the derivative evaluated at the point value $u_0$, we get

\begin{equation}
    \dfrac{d^2 x}{d\theta^2} + n^2_1 x = 0\,, \qquad  n^2_1 = (1 - J'(u_0))\,,
\end{equation}
which is the second-order harmonic oscillator equation. By integrating it, we obtain the solution
\begin{equation}
    x(\theta) = a_1 \, \cos (n_1 \theta + f_0) \, ,
\end{equation}
with arbitrary constant $f_0$ set equal to $f_0 = 0$. Compared to the closed Newtonian orbit $u = GM/h^2 \, (1 + \epsilon \cos\theta)$, if we define $\epsilon = h^2 a_1 / GM $, from Eq. (\ref{epi-perturbation}), we see that the motion of the particle is suitably described by $u = GM/h^2 \, (1 + \epsilon \cos n \theta)$, where $n$ leads to the orbital precession of relativistic origin. 

In fact, periastron occurs when the test particle arrives at the minimum distance point in the orbit given by the radius $r_0 = 1/u_0$, and corresponds to the maximum point of the variable $u$. This maximum point is reached when $\cos(n_1 \theta) = \cos (2\pi) = 1$, that is
$\cos \left( \sqrt{1 - J'(u_0)} \,\, \theta \right) = \cos (2\pi) = 1$, 
from which it follows 
\begin{equation}
\theta = 2\pi \, \left(\sqrt{1 - J'(u_0)}\right)^{-1}.
\end{equation}
\newpage
By expanding the Taylor series $(1-s)^{-1/2}$, $s = J'(u_0)$ and $s \ll 1$, it follows that 
\begin{equation}
    \theta \simeq 2\pi \, \biggl(1 + \dfrac{J'(u_0)}{2}\biggr)
\end{equation} and this quickly leads to the final quantity expressing the angular anomalistic precession of the total angle $\theta \simeq 2\pi + 2\pi \delta\theta \, = \, 2\pi + \Delta\theta$ wiped out by the test particle, that must be identified with the second term of the last relation as follows 
\begin{equation}
\Delta\theta_{ETG} \, = \, \pi J'(u_0) \, = \, - \dfrac{\pi}{h^2} V''_e (u_0)\,. 
\end{equation}
Therefore, performing a straightforward computation, we finally obtain 
\begin{equation}
    \Delta\theta_{ETG} \, = \, \Delta\theta_{GR} + \Delta\theta_p\qquad,\qquad \Delta\theta_{GR} = \frac{6 \pi GM}{a c^2 (1-\epsilon^2)}\,,
\end{equation}
where the General Relativity's contribution to the periastron advance stems from the first two terms at the second member of Eq. (\ref{Binet-equation}), and 
\begin{eqnarray}\label{additional-advance-in-ETG}
 \Delta\theta_{p} = - \dfrac{2\pi}{c^2} \, \Phi_{p}(u_0) - \dfrac{4 \pi u_0}{c^2} \, \Phi'_p (u_0) - \dfrac{\pi u^2_{0}}{c^2} \, \Phi''_p (u_0) - \dfrac{\pi}{h^2} \, \Phi''_p (u_0)\,,
\end{eqnarray}
represents the additional shift containing all the post-Newtonian corrections to the advance related to the corrective potentials coming from the theory (e.g. see Eqs. (\ref{Solutions-in-wfl-STFOG}), (\ref{STFOG-spherical-metric}), (\ref{NCSG-solutions}), (\ref{NCSG-spherical-metric})). The derivatives of the potentials are evaluated in $u_0 = [a(1-\epsilon^2)]^{-1}$. Putting all together, we find out
\begin{eqnarray}\label{periastron-advance-in-ETG}
    \Delta\theta_{ETG} \, = \, \frac{6 \pi GM}{a c^2 (1-\epsilon^2)} - \dfrac{2\pi}{c^2} \, \Phi_{p}(u_0) - \dfrac{4 \pi u_0}{c^2} \, \Phi'_p (u_0) - \dfrac{\pi u^2_{0}}{c^2} \, \Phi''_p (u_0) - \dfrac{\pi}{h^2} \, \Phi''_p (u_0) \, .
\end{eqnarray}

It must be noted that this final expression holds for small perturbations from the circular orbit and could be applied to systems where small eccentricities are involved, for example, as in the case of the Solar System.  

\subsection{Solution for large deviations from circularity}
However, if we want to include more general elliptic orbits whose deviations from circularity are remarkable, therefore involving much larger eccentricities as occurs in the Sgr A* cluster, we have to consider deviations so large that the Taylor series of $J(u)$ in Eq. (\ref{taylorseries1}) must be adequately expanded to the order \cite{Goldstein}
\begin{equation}\label{taylorseries2}
    J(u) = J(u_0) + J'(u_0) x + \dfrac{J''(u_0)}{2} x^2 + \dfrac{J'''(u_0)}{6} x^3 + O(u^4_0)\,.
\end{equation}

Referring to the identity $J(u) = h^{-2} V'(u)$, this implies that around the stable equilibrium point, the potential has even been developed up to the fourth order of the series, which can be physically considered a good approximation. Thus, the equation of the orbit assumes the new form
\begin{equation}\label{new-eq-orbit}
     \dfrac{d^2 x}{d\theta^2} + n^2_1 \, x = \dfrac{J''}{2} x^2 + \dfrac{J'''}{6} x^3 \,,
\end{equation}
with $ n^2_1 = 1 - J'(u_0)$. To solve this equation, we resort to the Poincaré-Lindstedt method \cite{poincare,Verhulst}. Hence, we now expand $x$ into a perturbative series as
\begin{equation}\label{expanded-sol}
    x \, = \, x_1 + x_2 + x_3 \,,
\end{equation}
with $ x_1 = a_1 \cos n \theta $, and thus perform the change of variable $\sigma = n \theta$, where
\begin{equation}\label{expanded-freq}
    n \, = \, n_1 + n_2 + n_3\,, 
\end{equation}
represents the exact value of the new total frequency with respect to the contributions due to the larger deviations from circular orbit. Therefore, the fundamental frequency $n_1$ undergoes a variation, with respect to its initial value, that is not affected by successive values. This is what we are looking for in achieving the final expression for the periastron precession determination. Eq. (\ref{new-eq-orbit}) can be rewritten as
\begin{equation}\label{new-eq-orbit2}
n\, \dfrac{d^2 x}{d\sigma^2} + n^2_1 \, x = \dfrac{J''}{2} x^2 + \dfrac{J'''}{6} x^3 \,.
\end{equation}

After inserting Eqs. (\ref{expanded-sol})-(\ref{expanded-freq}) in Eq. (\ref{new-eq-orbit2}), we obtain the following system of differential equations in the corresponding required order of expansion
\newpage
\begin{eqnarray}
\label{1de}
&& \dfrac{d^2 x_1}{d\sigma^2} + \, x_1 \, = \, 0\,,
\nonumber\\\nonumber\\
\label{2de}
 && \dfrac{d^2 x_2}{d\sigma^2} + \, x_2 \, = \, -\dfrac{2 n_2}{n_1} \dfrac{d^2 x_1}{d\sigma^2} + \dfrac{1}{2 n^2_1} J''\,x^2_1\,,
 \nonumber\\\nonumber\\
\label{3de}
&& \dfrac{d^2 x_3}{d\sigma^2} + \, x_3 \, = \, -\dfrac{2 n_3}{n_1} \dfrac{d^2 x_1}{d\sigma^2} -\dfrac{n^2_2}{n^2_1} \dfrac{d^2 x_1}{d\sigma^2} - \dfrac{2 n_2}{n_1} \dfrac{d^2 x_1}{d\sigma^2} 
 + \dfrac{1}{n^2_1} J'' \, x_1 x_2 + \dfrac{1}{6 n^2_1} J''' \, x^3_1\,.
\end{eqnarray}
From the homogeneous Eq. (\ref{1de}), we readily have 
\begin{equation}\label{sol-1de}
x_1(\sigma) = a_1 \cos (\sigma)\,,
\end{equation}
where the arbitrary constant is set to $\sigma_0 = 0$; if we substitute it into Eq. (\ref{2de}), it becomes
\begin{equation}\label{new-2de}
     \dfrac{d^2 x_2}{d\sigma^2} + \, x_2 \, = \, \dfrac{2n_1}{n_2} a_1 \cos{\sigma} +\dfrac{a_1 J''}{4 n^2_1} (1 + \cos{2\sigma})\,.
\end{equation}
We notice that the first term in the second member of the equation is a resonant term, since it has the same frequency of solution $x_1$; it has an increasing amplitude over time\footnote{This is physically incompatible with the system itself, in the absence of external sources of energy.} and would lead to a solution that includes an aperiodic secular term, responsible for unbounded growth with respect to $\sigma$. Since epycicles are almost periodic oscillations around the equilibrium point, we have to impose the condition for the vanishing of such a resonant factor. This leads us to set $n_2 = 0$. Then, by solving Eq. (\ref{new-2de}), the solution reads
\begin{equation}\label{sol-2de}
    x_2(\sigma) = a_2 \cos{\sigma} + a_2 \sin{\sigma} + \dfrac{a^2_1 J''}{4 n^2_1} \left( 1 - \dfrac{\cos{2\sigma}}{3} \right)\,.
\end{equation}
Now, by inserting Eqs. (\ref{sol-1de}) and (\ref{sol-2de}) in Eq. (\ref{3de}), we have
\begin{equation}
    \dfrac{d^2 x_3}{d\sigma^2} + \, x_3  =  \left( \dfrac{2n_3}{n_1} a_1 + \dfrac{5(J'')^2 }{24 n^4_1} a^3_1 + \dfrac{J'''}{8 n^2_1} a^3_1
 \right) \cos{\sigma} + \dfrac{J''}{2n^2_1} a_1 a_2 \left( 1 + \cos{2\sigma} + \dfrac{a_3}{a_2}\sin{2\sigma}  \right) + \dfrac{a^3_1}{24} \left( \dfrac{J'''}{n^2_1} - \dfrac{(J'')^2}{n^4_1} \right) \cos{3\sigma}\,,
\end{equation}
Applying, once again, the vanishing condition of the resonant factor on the $\cos{\sigma}$ term amplitude, we find a non-zero value for the correction $n_3$ to the fundamental frequency $n_1$ (as a relation between $n_1$ and $n_3$) of the form
\begin{equation}\label{frequency-correction}
    n_3 \, = \, -  5\dfrac{(J'')^2}{48 n^3_1} - \dfrac{J'''}{16 n_1} a^2_1 \,.
\end{equation}
and the consequent solution turns out to be
\begin{equation}\label{sol-3de}
    x_3(\sigma) = a_3 \cos{\sigma} + a_3 \sin{\sigma} + a_1 a_2 \dfrac{J''}{2 n^2_1} \left( 1 - \dfrac{\cos{2\sigma}}{3} - \dfrac{a_3 \sin{2\sigma}}{3 a_2} \right) + \dfrac{((J'')^2 - n^2_1 J''')}{192 n^4_1} \cos{3\sigma}\,.
\end{equation}
We recall that the $a_2$ amplitude of the solution (\ref{sol-2de}) has a magnitude smaller than $a_1$, and the $a_3$ amplitude of the solution (\ref{sol-3de}) must be of even lower order than $a_2$.

Finally, repeating the procedure in Section 3.2., the test particle arrives at the periastron when the maximum point of the orbit $u$ is reached, that is, when $ \cos{\sigma} = \cos{n \theta} = \cos{2\pi}$. Here, $n$ is the new value that contains corrections to the fundamental frequency $n_1$; therefore, we insert this expression, together with Eq. (\ref{frequency-correction}), in Eq. (\ref{expanded-freq}), and obtain
\begin{equation}
    n = n_1 + n_2 + n_3 \quad \Rightarrow \quad n = (1-J'(u_0))^{1/2} -  \dfrac{5 a^2_1 \, J''(u_0)^2}{48 (1-J'(u_0))^{3/2}} + \dfrac{3 a^2_1 \, J'''(u_0)}{48 (1-J'(u_0))^{1/2}} \, .
\end{equation}
Thus, from the relation
\begin{equation}
    \cos{(n_1 + n_2 + n_2) \, \theta } = \cos{2 \pi}\,,  
\end{equation}
it follows 
\begin{equation}
    \theta = \dfrac{2\pi}{(1-J'(u_0))^{1/2} - \frac{1}{48}\left( 5\, a^2_1\, J''(u_0)^2 \, (1-J'(u_0))^{-3/2}  + 3 a^2_1 \, J'''(u_0) \,  (1-J'(u_0))^{-1/2} \right)} \, .
\end{equation}
Proceeding as before, a multivariable Taylor expansion series up to the first order provides the following\footnote{We recall that in the previous subsection, we defined the relation $\epsilon = h^2 a_1 / GM$, with $u_0=GM/h^2$.}
\begin{equation}
      \theta \simeq 2\pi \, \biggl(1 + \dfrac{J'(u_0)}{2} + \dfrac{a^2_1}{16} J'''(u_0) + \dfrac{3 a^2_1}{32} J'(u_0) \, J'''(u_0) \biggr)\,,
\end{equation}
from which, we quickly infer that the total precession of the periastron with respect to the angle $\theta = 2\pi + \Delta\theta$ wiped out by the test particle, is given by 
\begin{equation}
    \Delta\theta_{ETG} = \pi J'(u_0) + \dfrac{1}{8} J'''(u_0) \, (u_0\,\epsilon)^2 +   \dfrac{3}{16} J'(u_0)\, J'''(u_0) (u_0\,\epsilon)^2 \,. 
\end{equation}
Therefore, recalling Eq. (\ref{J-function}) and discarding all the negligible higher order terms, we finally find
\begin{eqnarray}
\label{periastron-advance-in-ETG2}
&& \Delta\theta_{ETG} \, = \, \Delta\theta_{GR} + \Delta\theta_{p}\qquad , \qquad \Delta\theta_{GR} \, = \, \dfrac{6\pi GM}{a c^2 (1-\epsilon^2)}\,;
\end{eqnarray}
\begin{eqnarray}
\label{additional-advance-in-ETG2}
  \Delta\theta_{p} = &&  - \dfrac{2\pi}{c^2} \, \Phi_{p}(u_0) - \dfrac{4 \pi u_0}{c^2} \, \Phi'_p (u_0) - \dfrac{\pi u^2_{0}}{c^2} \, \Phi''_p (u_0) - \dfrac{\pi}{h^2} \, \Phi''_p (u_0) + \,
 \nonumber\\\nonumber\\
 && \qquad\,\, - \dfrac{3 \pi u^2_{0} \, \epsilon^2}{2 c^2} \, \Phi''_p (u_0) - \dfrac{\pi u^3_{0} \, \epsilon^2}{c^2}  \, \Phi'''_p (u_0) -  \dfrac{\pi u^2_{0} \, \epsilon^2}{8 h^2}  \, \Phi^{(4)}_p (u_0)\,.
\end{eqnarray}

This solution is entirely analytical, and the determination of the relativistic periastron advance beyond Einstein theory is now simply resorted to this final formula, including the post-Newtonian terms of the perturbing potential and also holding for large orbital eccentricities due to remarkable deviations from the circularity. In fact, thanks to Eqs. (\ref{periastron-advance-in-ETG2})-(\ref{additional-advance-in-ETG2}), the corrective and total precession are easily calculated, respectively. 
By comparison, we notice that the latter formula yields an extension of Eq. (\ref{additional-advance-in-ETG}) which is valid for binary systems where eccentricities are small (e.g. Solar System), but does not rely on the best physical approximation when applied to orbits with higher eccentricities (for instance, the S2 star around the Sgr A* black hole). Especially Eq. (\ref{additional-advance-in-ETG2}) consistently reduces to (\ref{additional-advance-in-ETG}) for low eccentricities of the elliptic orbit, since the other four successive contributions become negligible. Hence, the pair of equations (\ref{periastron-advance-in-ETG2})-(\ref{additional-advance-in-ETG2}) are generally much more indicated for gravitational tests where a high level of accuracy is required, for example, beyond the Solar System (whose planetary orbits have low eccentricities). 

This relativistic precession formula has universal validity independently of a given class of theories, according to the assumptions of radial symmetry of the model, consistency with GR, and Lagrangian $2L=g_{\mu\nu}\dot{x}^\mu\dot{x}^\nu$, without the need to choose a specific method, which can be convenient only if used for a certain theory. Eqs. (\ref{periastron-advance-in-ETG2})-(\ref{additional-advance-in-ETG2}) are then an effective product of the epiciclyc method and reduce the evaluation of the periastron advance to a direct application of the analytic formula (\ref{periastron-advance-in-ETG}), independently of the analytic form of the perturbing potential (Yukawa, power-law or logarithmic) and its nature. 
Furthermore, it turns out to be economical because it allows for a fast and simple calculation, since derivatives are much easier to compute than integrals. It does not require numerical integration techniques if the analytic form of the potential is too laborious or even impossible to treat when a given method is employed. In particular, it comprises all post-Newtonian terms at the required level of accuracy, thus enabling improvements by several orders of magnitude of the previous bounds on the theories. Alternatively, it can be easily used for testing gravitational effects if the physical parameters of a theory/model have already been estimated. 

In the end, it also provides an exact mathematical framework for constructing further orbital simulations without making use of numerical techniques or codes. In the following, we directly apply it in its more general form (\ref{additional-advance-in-ETG2}) in order to perform high-precision tests on the Solar System and the S2 star for the STFOG, NonCommutative Geometry and Quintessence Field.

\section{Tests on the Solar System}

\subsection{Applications to the ETG}
In this section, we apply the previous result and constrain the sizes of the hypotetical fifth forces arising from Scalar-Tensor-Fourth-Order Gravity, NonCommutative Geometry and Quintessence Field, respectively. The analysis is carried out by calculating the analytical expression in Eq. (\ref{periastron-advance-in-ETG2}) for their relativistic precession shifts. The Eqs. (\ref{ST_FOG_FE_NL_sol_ball}) show that the STFOG field equations lead to a gravitational potential of the Yukawa-like form ($r=|{\bf x}|$)
 \begin{equation}\label{VFOG2}
 \Phi(r)= - \frac{GM}{r} \left(1+\sum_{i=\pm, Y} F_i e^{-\beta_i r}\right)\,,
 \end{equation}
where $F_i$ and $\beta$ are the strength and range of the interaction corresponding to each mode $i=+, -, Y$. Referring to the ball-like solution for a non-rotating source\footnote{It means that the $g_{0i}$ mixed term of the metric is set to $0$. For example, in certain models like the one we are treating, it is a good assumption when the rotation of the source is so small that its influence can be neglected.} (\ref{Solutions-in-wfl-STFOG}) and the equations. (\ref{PhiFOG}) and (\ref{PsiFOG}), and comparing (\ref{VFOG2}) with the form of a Yukawa potential
\begin{equation}\label{Yukawapot}
V_Y(r) = \alpha \frac{e^{-r/\lambda}}{r}\equiv \alpha \frac{e^{-\beta r}}{r}\,,
\end{equation}
it follows the correspondence
\begin{equation}\label{alphaFOG}
 \alpha \to -\,GM F_i\,, \quad \beta \to \beta_i\,, \quad i=\pm, Y\,,
\end{equation}
with
\begin{equation}\label{FparameterFOG}
 F_+ =g(\xi,\eta)\,F(m_+ {\cal R})\,, \quad F_- = \Big[\frac{1}{3}-g(\xi,\eta)\Big]\,F(m_- {\cal R})\,, \quad
 F_Y=- \frac{4}{3}\,F(m_Y {\cal R})\,,
\end{equation}
\begin{equation}\label{betaFOG}
 \beta_\pm = m_R \sqrt{w_\pm} \,, \qquad \beta_Y = m_Y\,.
\end{equation}
We can now proceed with the analysis for the STFOG and NCSG.

\subsubsection{Scalar-Tensor-Fourth-Order Gravity}
\begin{figure}[!ht]
    \centering
    \subfigure[]{\includegraphics[width=0.45\textwidth]{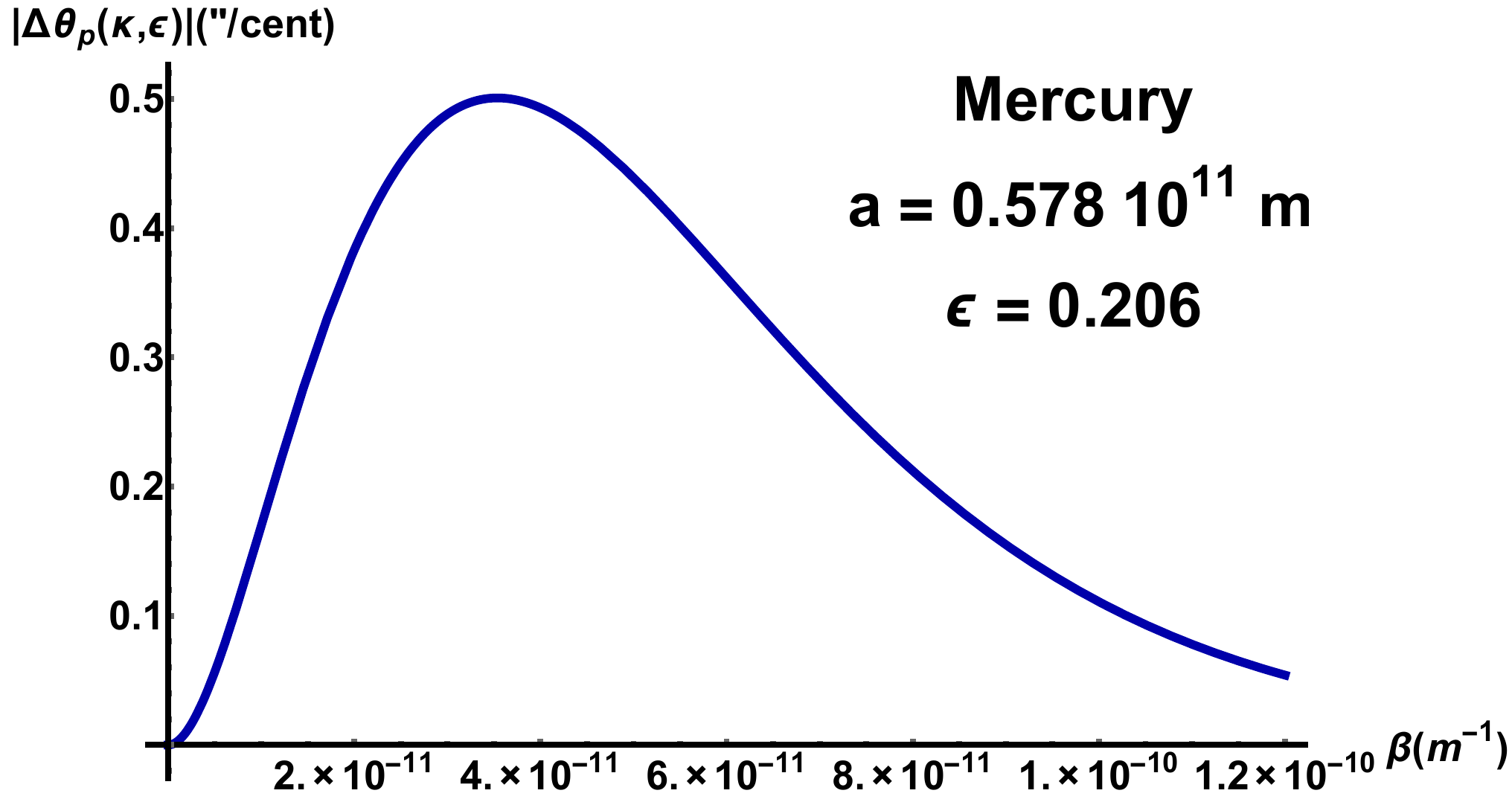}}
    \subfigure[]{\includegraphics[width=0.45\textwidth]{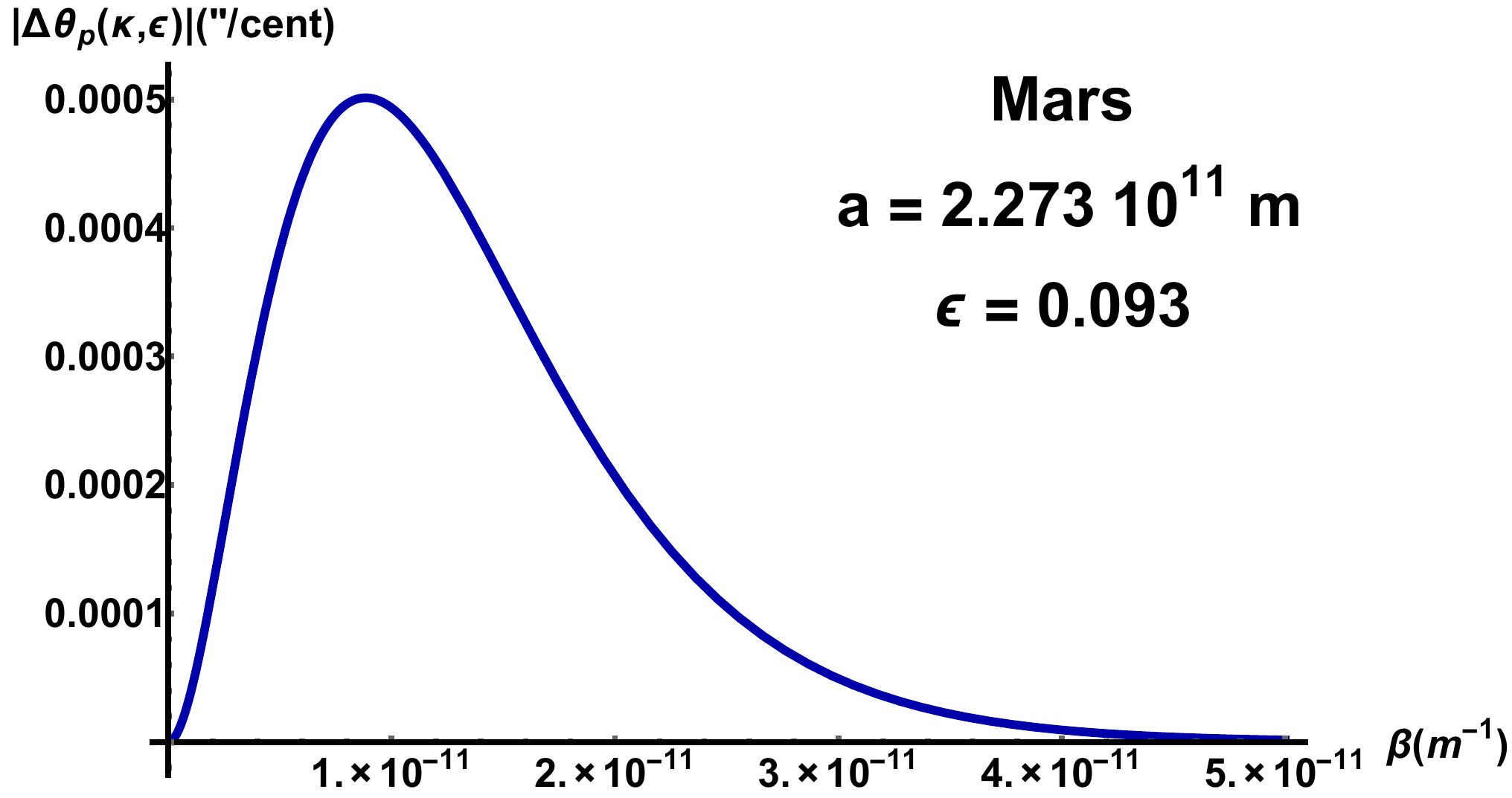}}
    \subfigure[]{\includegraphics[width=0.45\textwidth]{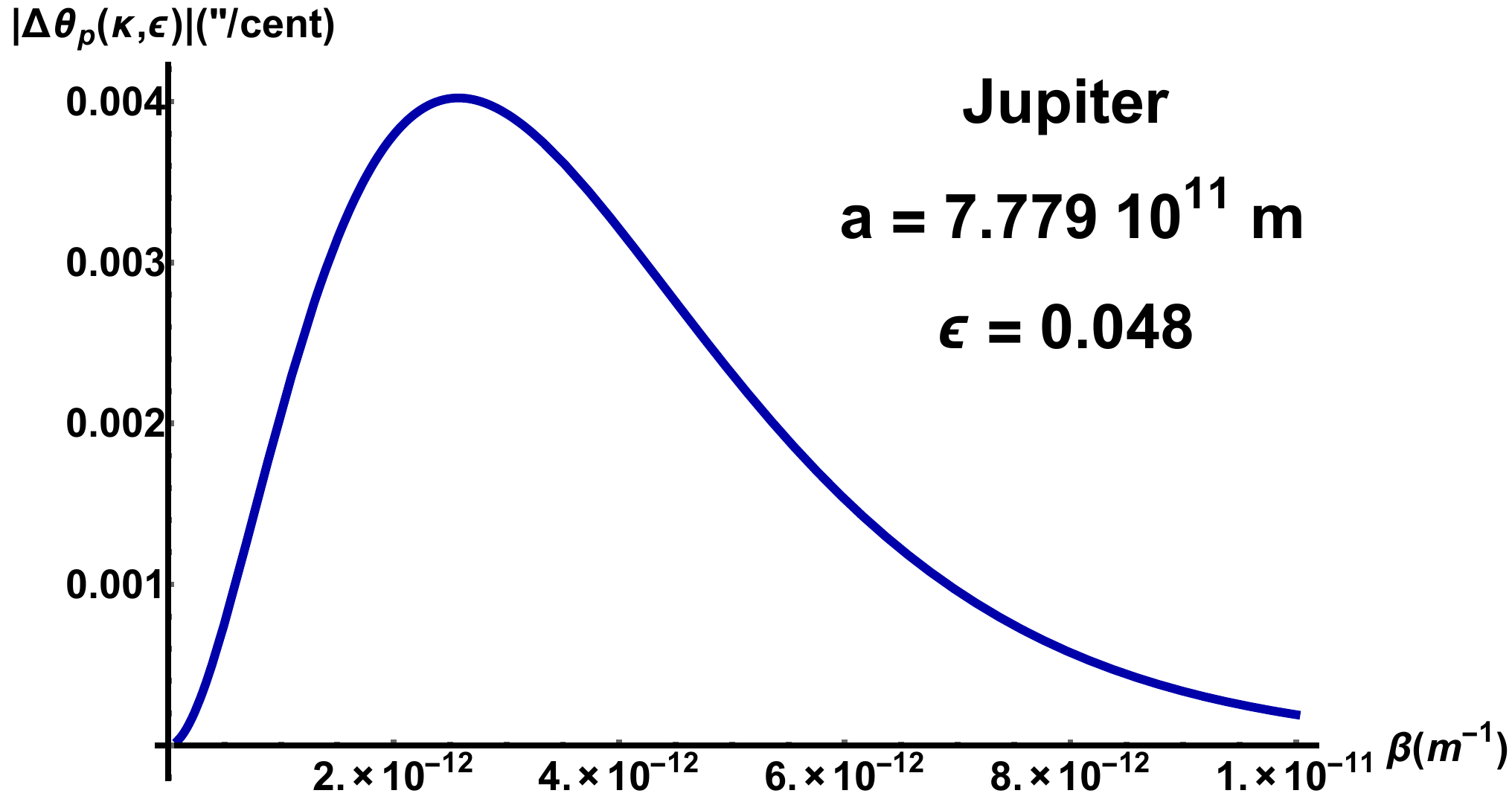}}
    \subfigure[]{\includegraphics[width=0.45\textwidth]{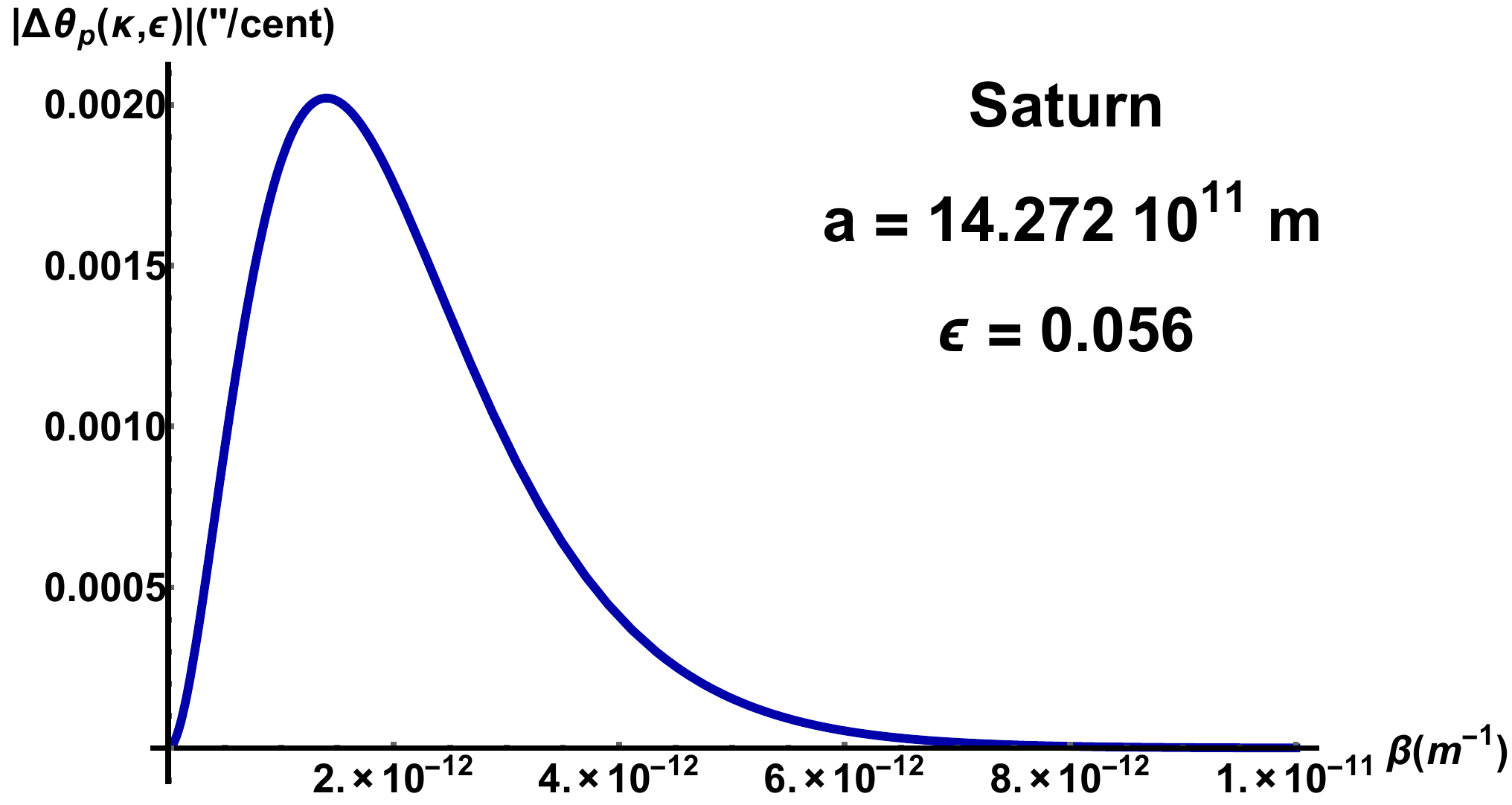}}
    \caption{(a) $|\Delta\theta_p(\kappa,\epsilon)|$ vs $\beta$ for Mercury. (b) $|\Delta\theta_p({\kappa,\epsilon})|$ vs $\beta$ for Mars. (c) $|\Delta\theta_p(\kappa,\epsilon)|$ vs $\beta$ for Jupiter. (d) $|\Delta\theta_p(\kappa,\epsilon)|$ vs $\beta$ for Saturn. $|\Delta \theta_p({\kappa,\epsilon})|$ is in $''/century$ units and is plotted by a blue line.}
    \label{fig:Newfoobar}
\end{figure}

On the basis of Eqs. (\ref{Solutions-in-wfl-STFOG}), (\ref{STFOG-spherical-metric}), (\ref{VFOG2}), (\ref{alphaFOG}), by applying formula (\ref{additional-advance-in-ETG2}), we determine the additional periastron advance due to the post-Newtonian terms is $\Delta \theta_p(\kappa,\epsilon)$ for the Scalar-Tensor-Fourth-Order Gravity, and find
\begin{eqnarray}\label{STFOG-periastron-advance}
 \Delta\theta_{p} ({\beta,\epsilon}) = && \sum_{i = \pm\,Y} F_i \, \biggl\{ \dfrac{6\pi GM}{ac^2(1-\epsilon^2)}\, +  \dfrac{4\pi GM}{c^2}\beta_i \,
     + \dfrac{\pi GM}{c^2} \beta^2_i a(1-\epsilon^2) \,  +  \pi \beta^2_i a^2(1-\epsilon^2)^2 + \dfrac{27\pi GM a \, \epsilon^2}{2c^2} (1-\epsilon^2)  +
     \nonumber\\\nonumber\\
     && + \dfrac{3 \pi \epsilon^2}{2} \beta^2  a^2 (1-\epsilon^2)^2 +  \dfrac{9\pi GM \,\epsilon^2}{c^2} \beta^2 a^2 (1-\epsilon^2)^2 + \dfrac{\pi GM\, \epsilon^2}{8 c^2} \beta^4  a^3  (1-\epsilon^2)^3 - \pi \beta^3 a^3 \epsilon^2 (1-\epsilon^2)^3 +
      \nonumber\\\nonumber\\
     && + \dfrac{9 \pi GM \epsilon^2}{8 c^2} \beta^4 a^3 (1-\epsilon^2)^3 + \dfrac{\pi\ \, \epsilon^2}{8} \beta^4 a^4 (1-\epsilon^2)^4 \biggr\} \, e^{-\beta_i a(1-\epsilon^2) }\,. 
\end{eqnarray}
Equation (\ref{periastron-advance-in-ETG2}) yields the total precession. We recall that $F_i$ and $\beta_i$ are the strength and range of the interaction corresponding to each mode $i=+, -, Y$ respectively, and their expressions are given in Eqs. (\ref{FparameterFOG}) and (\ref{betaFOG}). To infer theoretical constraints, we impose that the additional periastron shift $\Delta \theta_p(\kappa,\epsilon) $ given by (\ref{STFOG-periastron-advance}), with $\kappa = \beta a$, is less than the astrometric error $\eta$. Maximising $\Delta \theta_p(\kappa,\epsilon) $ with respect to the constrained problem
 \begin{equation}\label{New-Fibound}
 |\Delta \theta_p(\kappa,\epsilon)|\lesssim \eta \,,
 \end{equation}
 we obtain the bounds of the parameters $\{\beta,\,|F_i|\}$, with $i=\pm, Y$, by fixing a given known astrometric error $\eta$ and eccentricity $\epsilon$, where the maximum value of the precession $|\Delta \theta_p(\kappa,\epsilon)|$ is reached at the point $\beta_i = \beta^{max}_i$. In Fig. \ref{fig:Newfoobar}, the function $|\Delta\theta_p(\kappa,\epsilon)|$ is plotted relative to Mercury, Mars, Jupiter, and Saturn. In Table \ref{NewtableIIa}, the corresponding bounds on $F_i$ are reported, and, as we can see, the post-Newtonian contributions of relativistic origin allow us to achieve a further improvement on the bound of the theory.
\begin{table}[!ht]
\renewcommand{\arraystretch}{2}
\begin{center}
\caption{Values of periastron advance for the first six planets of the Solar System. In the table we present the values of the eccentricity $\epsilon$, semi-major axis $a$ in metres, the orbital period $P$ in years, the periastron advance predicted in General Relativity (GR).}\label{TabPlanets}
\begin{tabular}{c c c c c c }
	\Xhline{4\arrayrulewidth}
	Planet & $\qquad\epsilon\qquad$ & $\quad a \, (10^{11}\, m)\quad$ & $P\, (yrs)$ & $\quad \Delta\theta_{GR} \, (''/century)\quad $ & $\Delta\theta_{obs}$ \\
	  \hline
	Mercury &  $0.205$ & $0.578$  &  $0.24$ & $43.125$
	& $42.989   \pm 0.500$ \\
	
	Venus & $0.007$  & $1.077$ & $0.62$  &  $8.62$ &  $8.000\pm 5.000$   \\
	
	Earth & $0.017$  &  $1.496$ &  $1.00$ & $3.87$  &    $5.000\pm 1.000$  \\
	
	Mars  & $0.093$  &  $2.273$ & $1.88$  & $1.36$  &   $1.362\pm 0.0005$   \\
	
	Jupiter & $0.048$  &  $7.779$ & $11.86$   &  $0.0628$  & $0.070 \pm 0.004$  \\
	
	Saturn &  $0.056$ &   $14.272$ &  $29.46$  &   $0.0138$ &   $0.014 \pm 0.002$ \\
	\Xhline{4\arrayrulewidth}
\end{tabular}
\end{center}
\end{table}

\begin{table}[!ht]
\renewcommand{\arraystretch}{1.75}
\begin{center}
\caption{New bounds on $F_i$, $i=\pm, Y$ obtained from (\ref{New-Fibound})  using the values of periastron advance for planets of the Solar System.}
\label{NewtableIIa}
\begin{tabular}{c c c c }
	\Xhline{4\arrayrulewidth}
	Planet & $ \,\, |\eta|$ & $  \qquad\,\,\,\, \beta_i^{\text{max}}\,(m^{-1})\simeq $ & $  \quad\,\,\,\,\, |F_i|\lesssim$ \\
	  \hline
Mercury & $  \,\,\,\, 0.5$ & $   \qquad 3.54 \times 10^{-11} $ & $  \qquad 3.44 \times 10^{-12} $
\\
Mars & $  \quad\,\,\,\, 5 \times 10^{-4} $ & $  \qquad 8.84 \times 10^{-12} $ & $  \qquad 2.69 \times 10^{-11} $ 
\\
Jupiter & $  \quad\,\,\,\, 4\times 10^{-3} $ & $  \qquad 2.57 \times 10^{-12} $ & $  \qquad 1.36\times 10^{-9} $
\\
Saturn & $  \quad\,\,\,\, 2\times 10^{-3} $ & $  \qquad 1.40 \times 10^{-12} $ & $  \qquad 1.70 \times 10^{-9}$ \\
	\Xhline{4\arrayrulewidth}
\end{tabular}
\end{center}
\end{table}
\newpage
\subsubsection{NonCommutative Geometry}
\begin{figure}[!ht]
    \centering
    \subfigure[]{\includegraphics[width=0.45\textwidth]{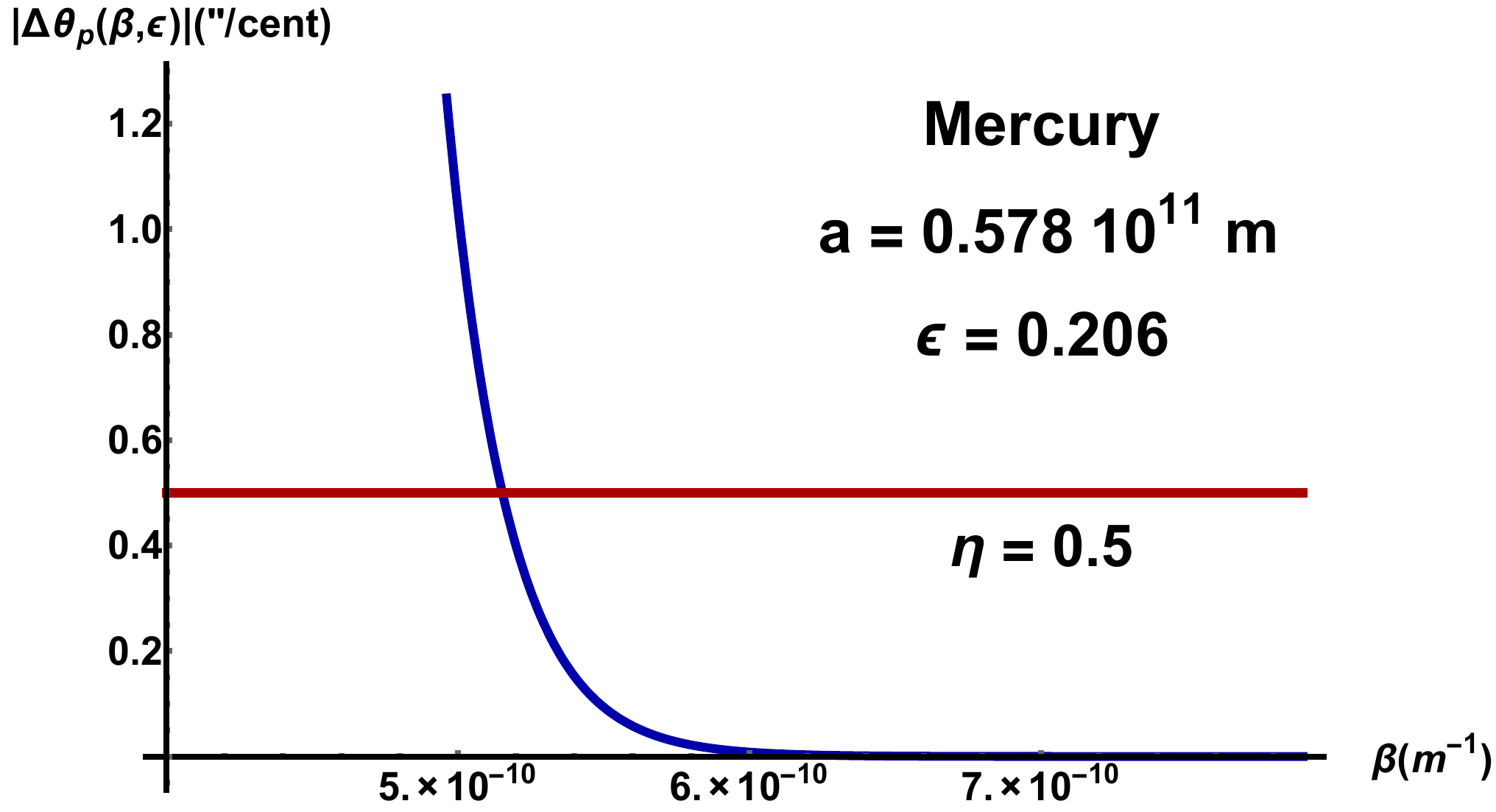}}
    \subfigure[]{\includegraphics[width=0.45\textwidth]{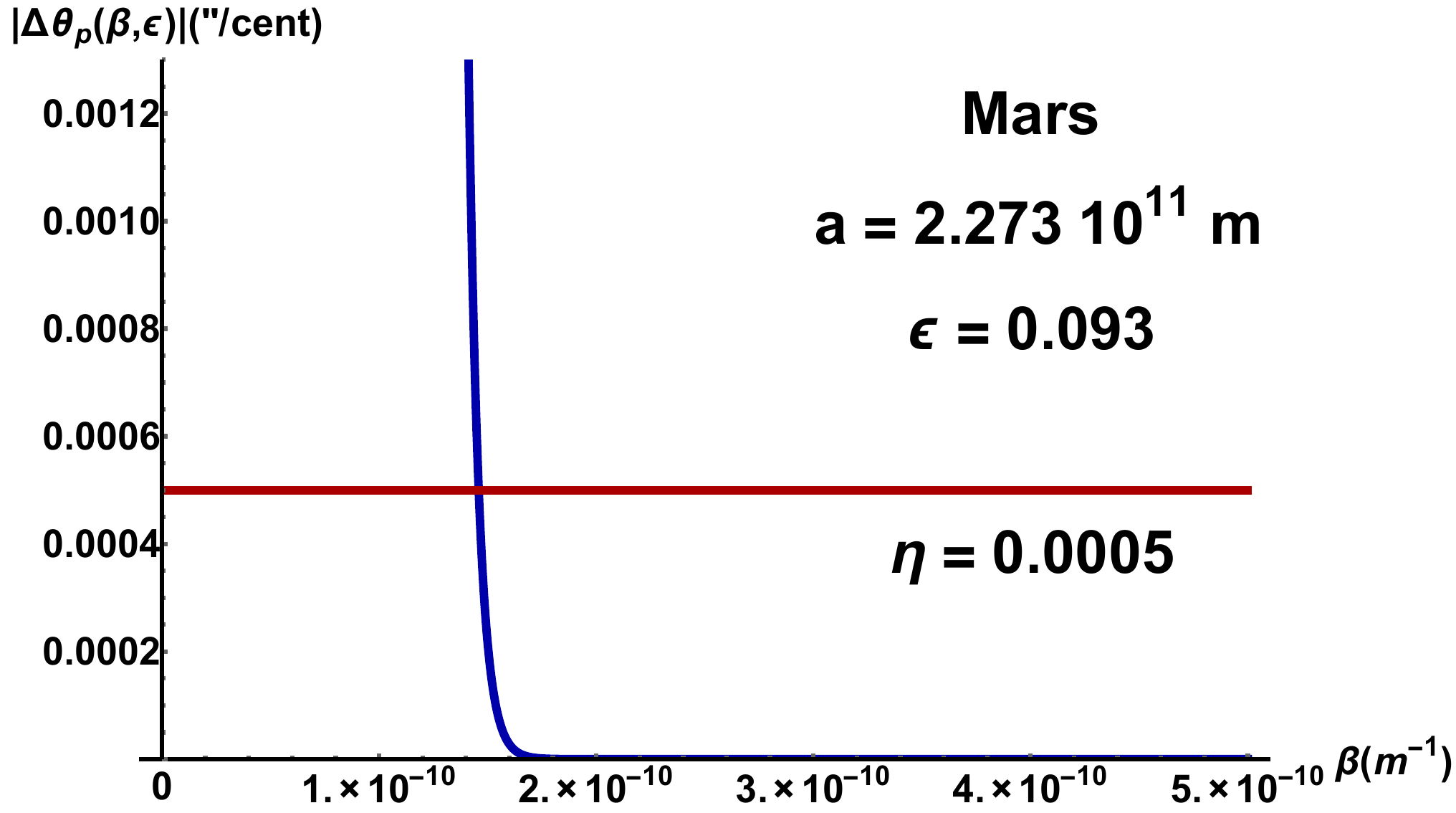}}
    \subfigure[]{\includegraphics[width=0.45\textwidth]{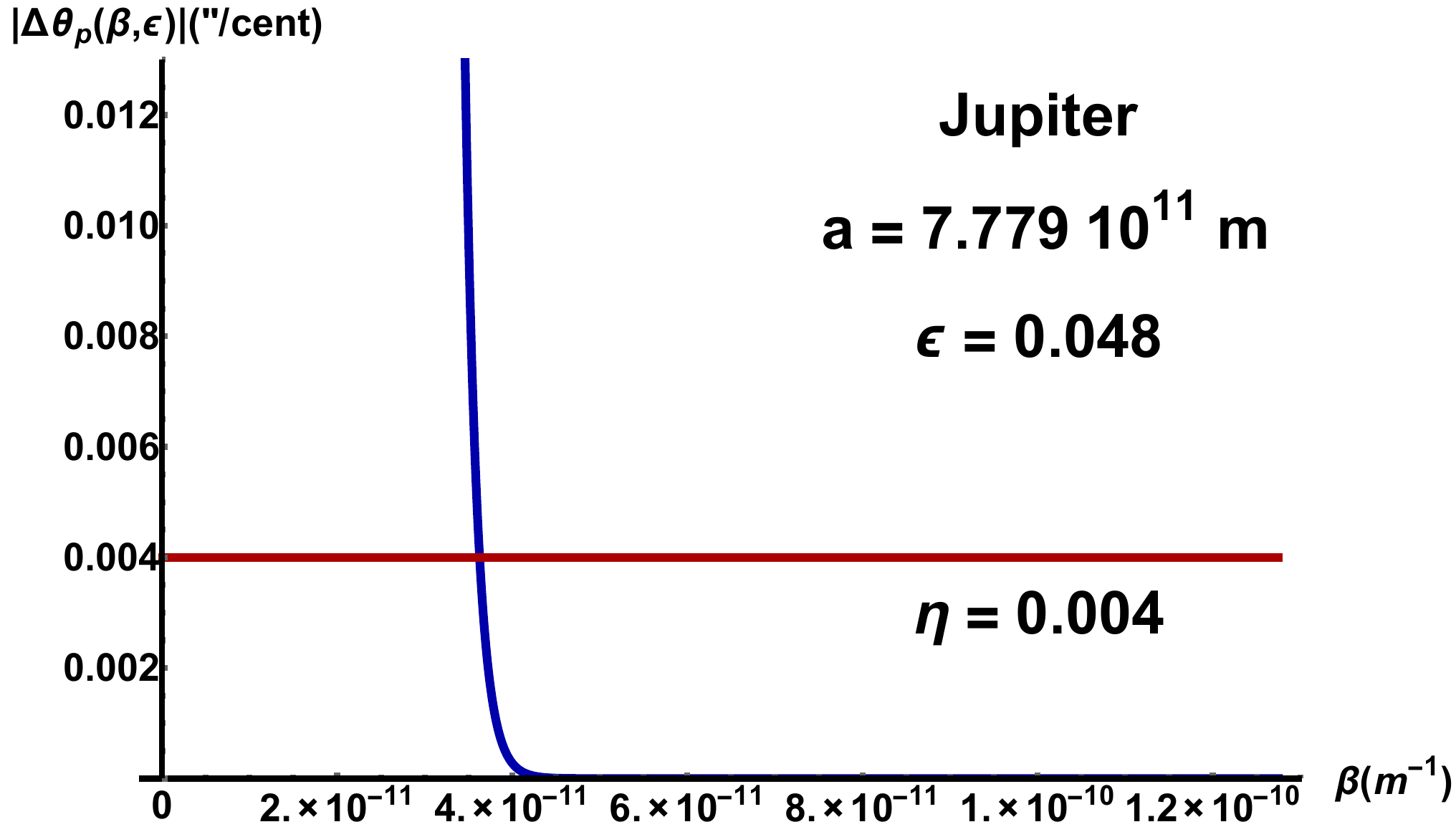}}
    \subfigure[]{\includegraphics[width=0.45\textwidth]{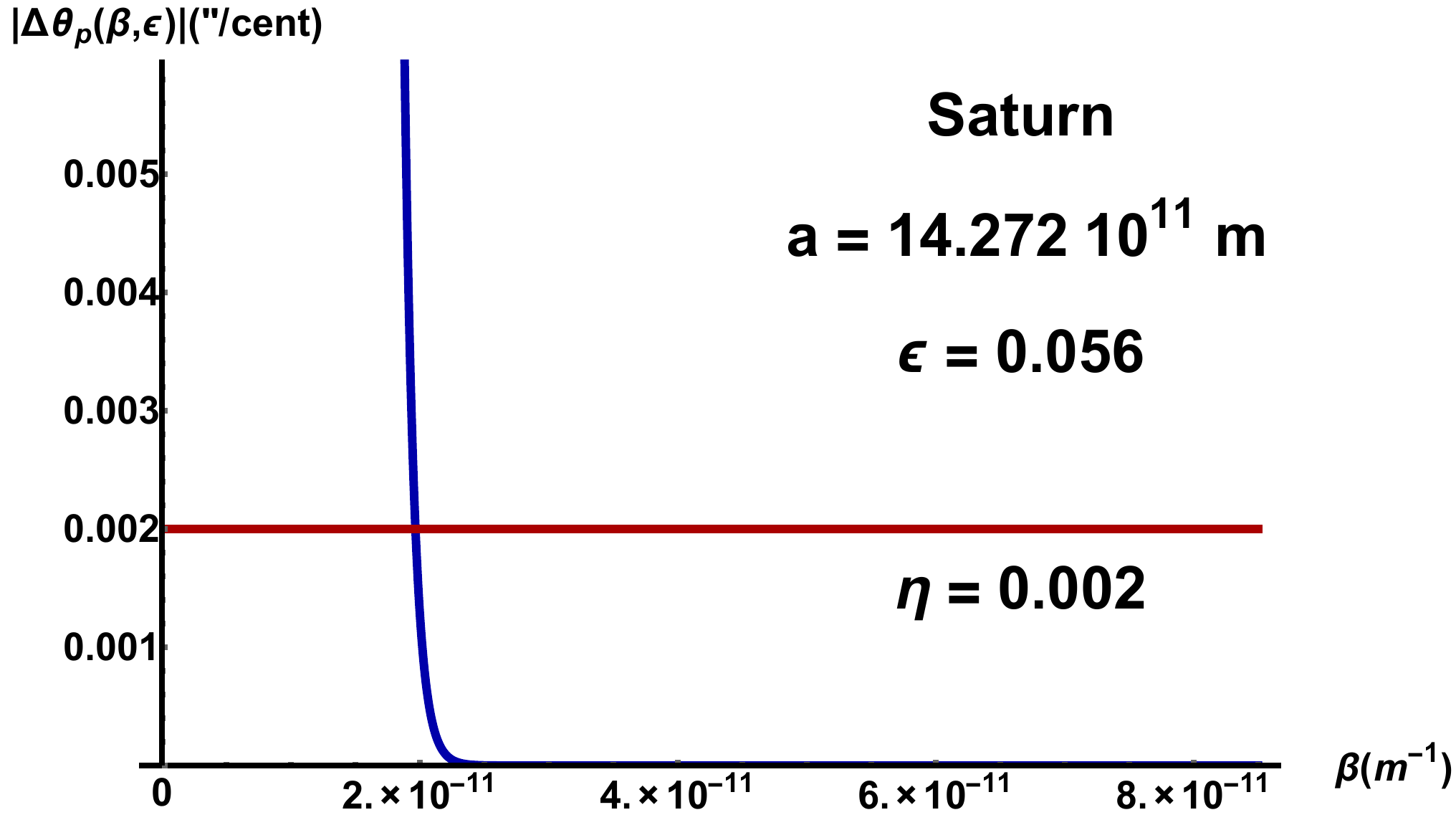}}
    \caption{(a) $|\Delta\theta_p({\beta,\epsilon})|$ vs $\beta$ for Mercury. (b) $|\Delta\theta_p({\beta,\epsilon})|$ vs $\beta$ for Mars. (c) $|\Delta\theta_p({\beta,\epsilon})|$ vs $\beta$ for Jupiter. (d) $|\Delta\theta_p({\beta,\epsilon})|$ vs $\beta$ for Saturn. $|\Delta \theta_p({\beta,\epsilon})|$ (blue line) and the constant $\eta$ (red line) are plotted in $''/century$ units.}
    \label{fig:NewNCSGfoobar}
\end{figure}

Considering the potential in Eqs. (\ref{NCSG-solutions}) and (\ref{NCSG-spherical-metric}), with the application of Eq. (\ref{additional-advance-in-ETG2}) we compute the additional post-Newtonian periastron advance, and find
\begin{eqnarray}\label{NCSG-periastron-advance}
     \Delta\theta_{p} ({\beta,\epsilon})  = && - \biggl\{ \dfrac{8\pi GM}{ac^2(1-\epsilon^2)} \, + \dfrac{16\pi GM}{3c^2}\beta 
     \,  + \dfrac{4\pi GM}{3c^2} \beta^2 a(1-\epsilon^2) \, + \dfrac{4\pi}{3} \beta^2 a^2(1-\epsilon^2)^2  + \dfrac{18 \pi GM \, \epsilon^2}{c^2}\beta^2 a (1 - \epsilon^2)
     \nonumber\\\nonumber\\
     && \, +  2 \pi \, \epsilon^2 \, \beta^2 a^2 (1 - \epsilon^2)^2 - \dfrac{ 12 \pi GM\, \epsilon^2}{c^2} \beta^3 a (1-\epsilon^2)^2 + \dfrac{6\pi GM \, \epsilon^2}{c^2} \beta^2 a^3 (1 - \epsilon^2)^3 - \dfrac{ 4\pi\, \epsilon^2}{3} \beta^3 a^3 (1-\epsilon^2)^3 \nonumber\\\nonumber\\
     && \, + \dfrac{3\pi GM \, \epsilon^2}{2} \beta^4 a^3 (1 -\epsilon^2)^3 + \dfrac{\pi \, \epsilon}{6} \beta^4 a^4 (1-\epsilon^2)^4 
     \biggr\} \, e^{-\beta  a(1-\epsilon^2)} \,.
\end{eqnarray}
Here, with respect to the adopted sign convention, the coupling constant of the induced Yukawa-like potential and the rage of interaction are 
\begin{equation}
\alpha = \dfrac{4}{3}GM\,,\qquad \beta = \beta_{NCSG}\,,
\end{equation}
respectively. The constraint in NCSG for the planets of the Solar System is identified by 
\begin{equation}\label{NewbetaNCSGbound}
    |\Delta\theta_p({\beta,\epsilon})| \lesssim \eta \quad \to \quad |\beta|\lesssim \,\widetilde{\Theta}(\eta,\epsilon) \,,
\end{equation}
where $\widetilde{\Theta}(\eta,\epsilon)$ is defined as the expression from which we infer the new bounds on $\beta$ with respect to a certain known value of the astrometric error $|\eta|$ and the eccentricity, or equivalently an upper bound on its characteristic length $\beta^{-1}$. Results are reported in Table \ref{NewtableNCG} (see also Fig. \ref{fig:NewNCSGfoobar}). These results show that the bounds on $\beta$ reach a further improvement in their precision
$\beta \geq 7.55\times 10^{-13} \, {\rm m}^{-1}$ \cite{Nelson:2010rt, Nelson:2010ru}.

\begin{table}[!ht]
\renewcommand{\arraystretch}{1.75}
\begin{center}
\caption{Lower bounds on $\beta$ obtained from (\ref{NewbetaNCSGbound}) using the values of periastron advance for the planets of the Solar System. }
\label{NewtableNCG}
\begin{tabular}{c c c}
	\Xhline{4\arrayrulewidth}
	Planet & $\quad\quad \,\, |\eta| \,\, $ &  $ \quad\quad\,\,\beta (m^{-1}) > \,\,$ \\
	  \hline
Mercury & $\quad\quad 0.5$  & $\quad\quad 5.15 \times 10^{-10}  $\\
Mars & $\quad\quad\quad 5 \times 10^{-4}$  & $\quad\quad 1.46 \times 10^{-10}$ \\
Jupiter & $\quad\quad\quad 4\times 10^{-3}$ & $\quad\quad 3.61 \times 10^{-11}$ \\
Saturn & $\quad\quad\quad  2\times 10^{-3}$ & $\quad\quad 1.97 \times 10^{-11}$ \\
	\Xhline{4\arrayrulewidth}
\end{tabular}
\end{center}
\end{table}

\subsubsection{Quintessence Field}
\begin{figure}[!ht]
    \centering
    \subfigure[]{\includegraphics[width=0.45\textwidth]{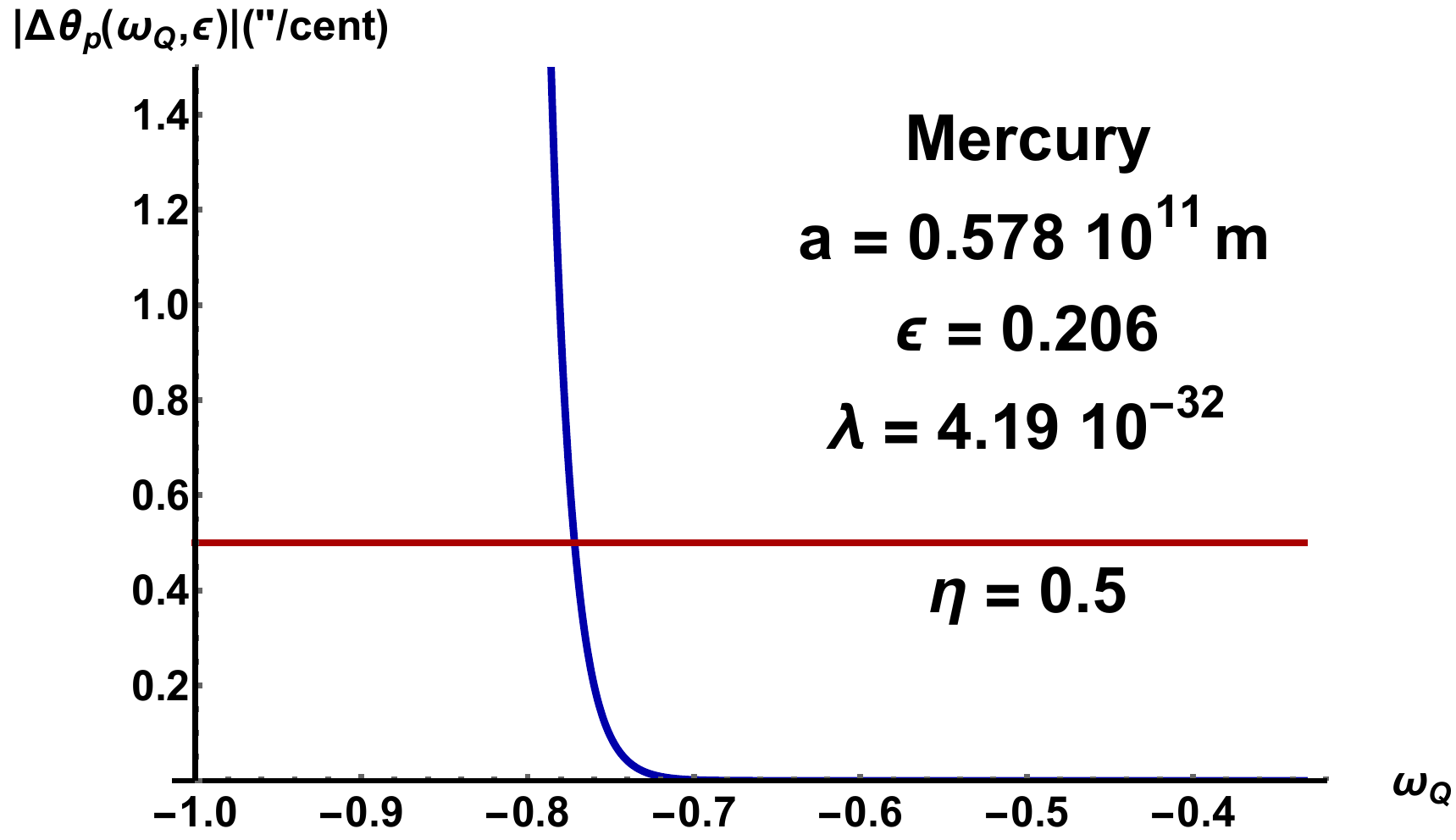}}
    \subfigure[]{\includegraphics[width=0.45\textwidth]{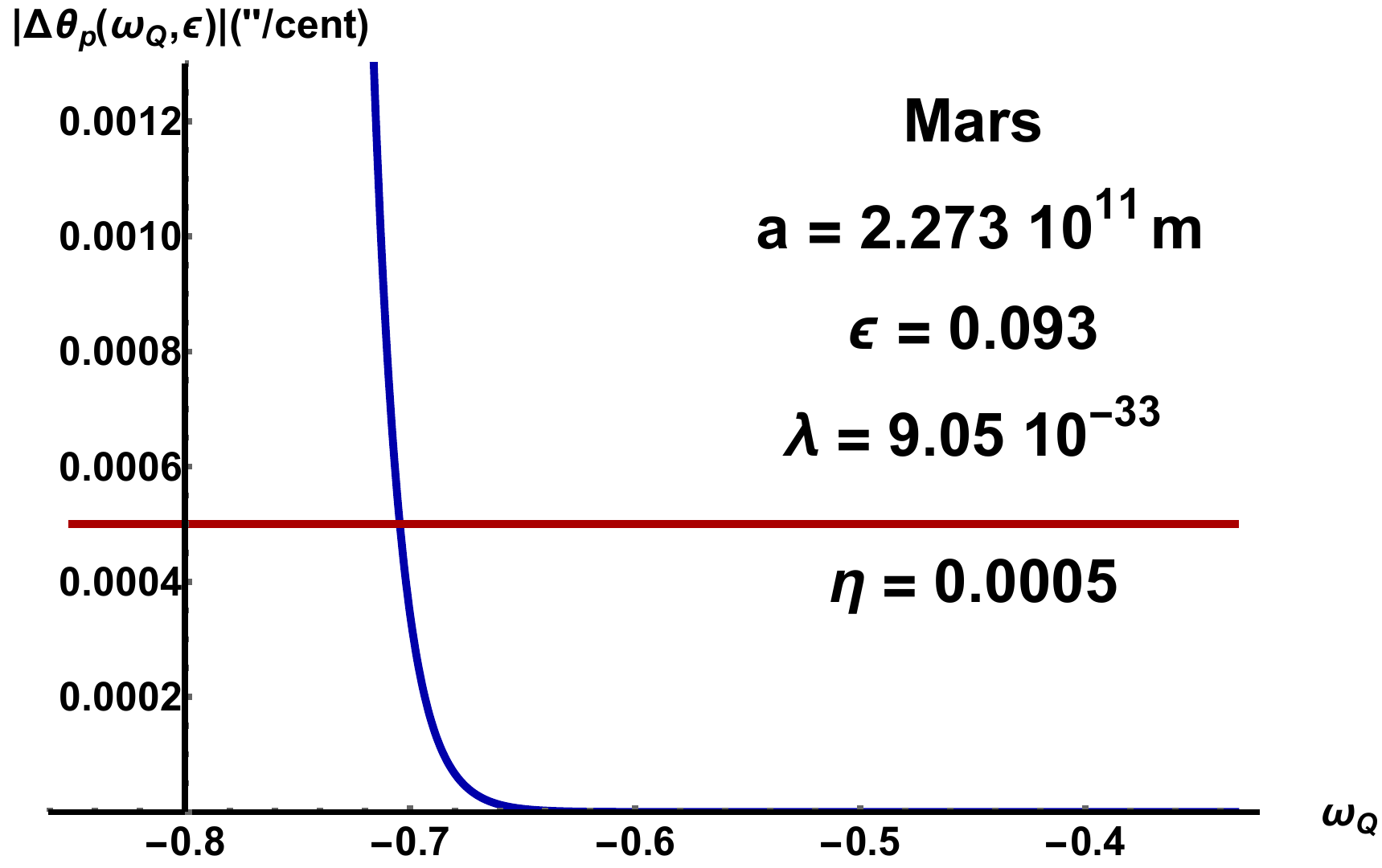}}
    \subfigure[]{\includegraphics[width=0.45\textwidth]{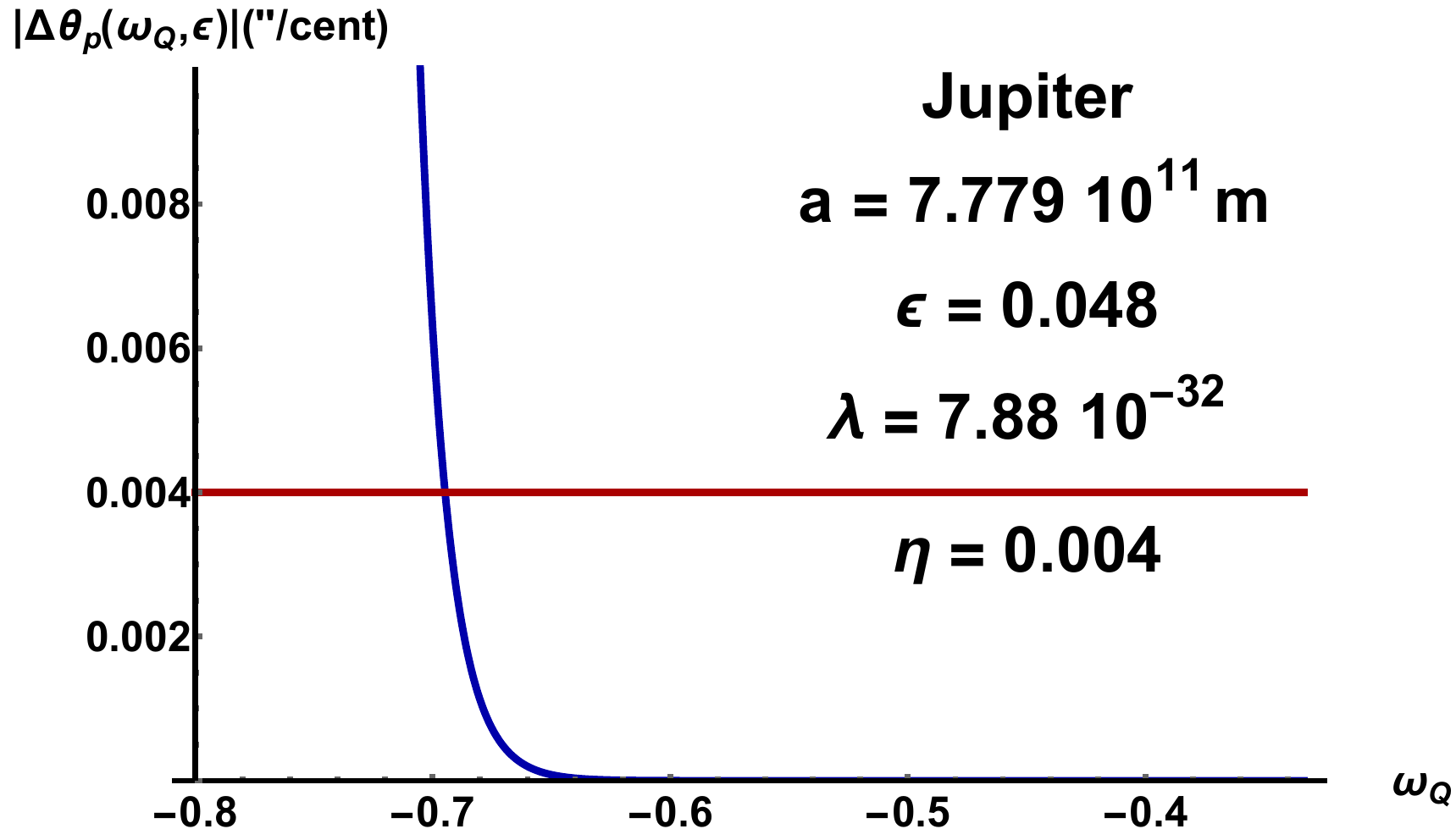}}
    \subfigure[]{\includegraphics[width=0.45\textwidth]{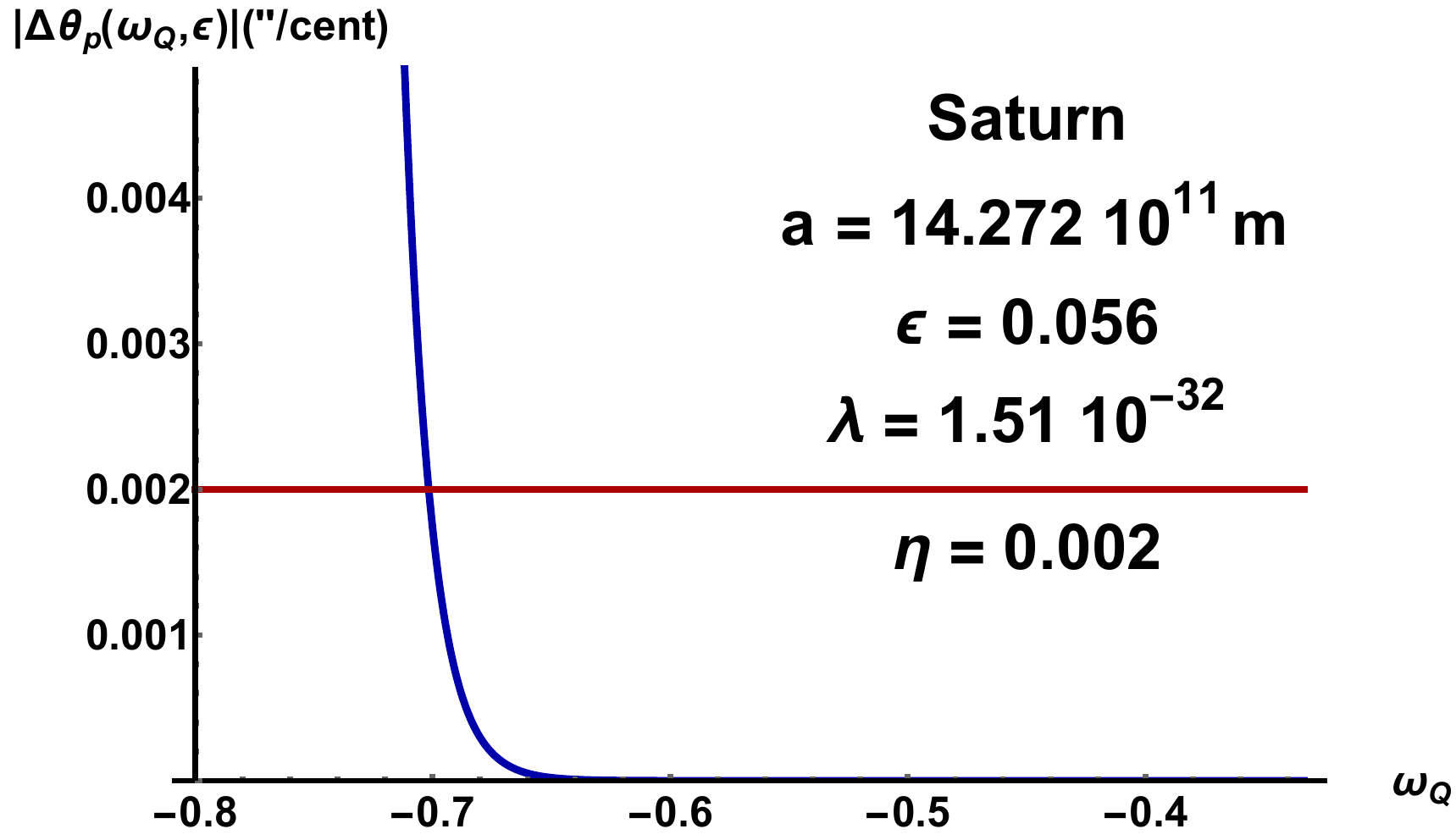}}
    \caption{(a) $|\Delta \theta_p(\omega_Q, \epsilon)|$ vs $\omega_Q$ for Mercury. (b) $|\Delta \theta_p(\omega_Q, \epsilon)|$ vs $\omega_Q$ for Mars. (c) $|\Delta \theta_p(\omega_Q, \epsilon)|$ vs $\omega_Q$ for Jupiter. (d) $|\Delta \theta_p(\omega_Q, \epsilon)|$ vs $\omega_Q$ for Saturn. $|\Delta \theta_p(\omega_Q, \epsilon)|$ (blue line) and the constant $\eta$ (red line) are plotted in $''/century$ units. Coloured gray zone indicates values outside the range $-1 \leq\omega_Q\leq -1/3$.}
    \label{figPowLaw}
\end{figure}

The Quintessential Potential reads $\Phi_p(r)=-\frac{\lambda}{r^{3\omega_Q+1}}$, so that compared to a power-law potential $V_{PL} (r)\, = \, \alpha_q \, r^q$ and Eq. (\ref{frmetric}), one has
 \begin{equation}
 q\to -(3\omega_Q+1)\qquad \alpha_q \to \lambda\,.
 \end{equation}
Concerning the additional relativistic precession (\ref{additional-advance-in-ETG2}) due to the Quintessence Field, we obtain the expression
\begin{eqnarray} \label{precesspowerlaw2}
\Delta \theta_p(\omega_Q, \epsilon) = && \dfrac{2\pi \lambda}{[a(1-\epsilon^2)]^{3\omega_Q + 1}} + \dfrac{4\pi \lambda (3\omega_Q + 1)}{[a(1-\epsilon^2)]^{3\omega_Q + 1}} + \dfrac{3\pi \omega_Q \lambda (3\omega_Q + 1)}{[a(1-\epsilon^2)]^{3\omega_Q + 1}}  - \dfrac{6 \pi \omega_Q \lambda \, (3\omega_Q + 1)}{ r_s [a(1-\epsilon^2)]^{3\omega_Q}} +  
\nonumber\\\nonumber\\
&& \qquad \,\, + \dfrac{ 9 \pi \omega_Q \lambda (3\omega_Q + 1) \epsilon^2}{2[a(1-\epsilon^2)]^{3\omega_Q + 1}} + \dfrac{ 3 \pi \omega_Q \lambda (9\omega^2_Q - 1) \epsilon^2}{[a(1-\epsilon^2)]^{3\omega_Q + 1}} +  \dfrac{  3\pi \omega_Q \lambda (9\omega^2_Q - 1)(3\omega_Q - 2) \epsilon^2}{4 r_s [a(1-\epsilon^2)]^{3\omega_Q}} \,,
\end{eqnarray}
where the Schwarzshild radius is $r_s=2 GM/c^2.$ By requiring $|\Delta \theta_p(\omega_Q, \epsilon)| \lesssim \eta$ as a constrained problem, one gets the bounds of the parameters $\{\omega_Q, \lambda\}$. The results are reported in Table \ref{tablePowLaw} and Fig. \ref{figPowLaw} for the different values of $ \lambda $.

\begin{table}[!th]
\renewcommand{\arraystretch}{1.75}
\begin{center}
\caption{Values of the parameter $\omega_Q$ obtained from (\ref{precesspowerlaw2}) using the values of periastron advance for planets of the Solar System.}
\label{tablePowLaw}
\begin{tabular}{ c c c c }
	\Xhline{4\arrayrulewidth}
	Planet & $\quad\quad |\eta|$ &  $\quad\quad \lambda (m^{3\omega_Q+1})$ & $\quad\quad \omega_Q\gtrsim$ \\
	  \hline
Mercury & $\quad\quad\, 0.5$ & $\quad\quad 4.19\times 10^{-32}$ & $\quad\quad -0.78$ \\
Mars & $\quad\quad\quad\,  5 \times 10^{-4}$ & $ \quad\quad  9.05 \times 10^{-33}$ & $\quad\quad -0.71$ \\
Jupiter & $\quad\quad\quad\, 4\times 10^{-3}$ & $\quad\quad  7.88 \times 10^{-32}$ & $\quad\quad -0.69$ \\
Saturn & $\quad\quad\quad\, 2\times 10^{-3}$ & $\quad\quad  1.51 \times 10^{-32}$ & $\quad\quad -0.70$ \\
	\Xhline{4\arrayrulewidth}
\end{tabular}
\end{center}
\end{table}
\newpage
\section{Tests and orbital simulations on S2 star}
\begin{figure}[!ht]
    \centering
    {\includegraphics[width=0.7665\textwidth]{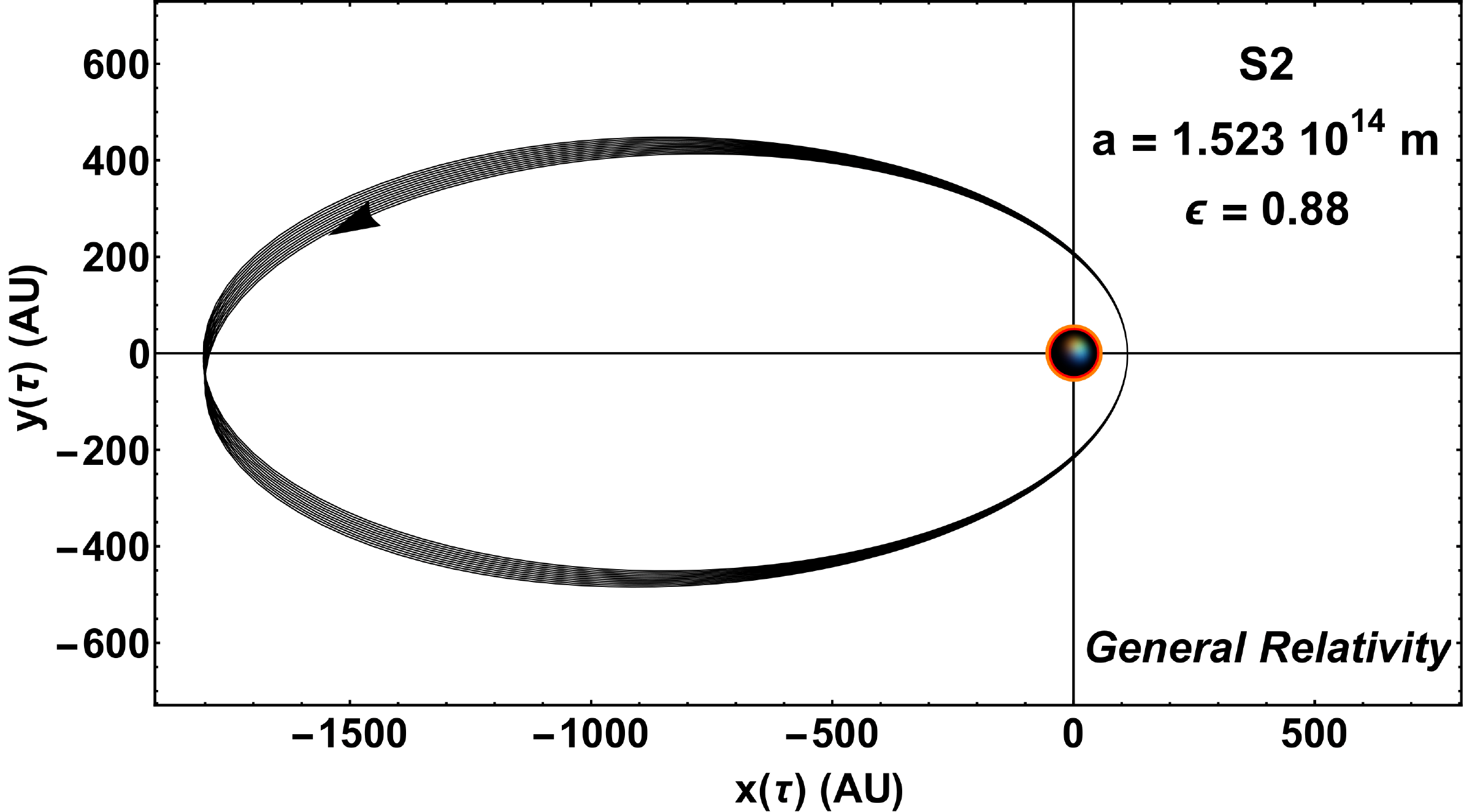}}
    \caption{Orbital simulation of the S2 star around Sgr A* Super Massive Black Hole obtained by numerical solution of the Eq. [\ref{Binet-equation}] for the GR case (recovered when $\Phi_p(r)=0$). Axes are measured in astronomical units (AU). Sgr A* is oversized for better display.}
    \label{orbsimGR}
\end{figure}

In this last section, we conclude our analysis testing the Extended Gravity predictions for S2 star orbiting around Sagittarius A*, the Super Massive Black Hole at the centre of the Milky Way. Sgr A * has a mass equal to $M=(4.5 \pm 0.6)\times 10^{6} M_{\odot}$  and a Schwarzschild radius $r_s = 2GM/c^2 = 1.27 \times 10^{10} \, \text{m}$, the eccentricity of its orbit is $\epsilon=0.88$, and the semimajor axis has a value $a=1.52917 \times 10^{14} \, \text{m}$. According to Ref. \cite{Abuter} (see also \cite{Will,Will2}), periastron precession of the S2 star is $0.2^{\,\,+0.057}_{\,-0.014}$ deg, therefore, a positive error $\eta=0.057$ and a negative error $\eta=-0.14$ with respect to the measured angle of precession. The general relativistic S2 star orbit around Sgr A* is reported in Fig. \ref{orbsimGR}, as an outcome of a numerical simulation carried out in order to illustrate the predicted GR's precession. In fact, the effect on dynamics is more clearly visible because of the large value of the orbital eccentricity and the proximity of S2 to the Black Hole, especially when it reaches the periastron.  

We now present the results on the periastron advance for the examined gravitational models. Orbital simulations have been numerically performed with respect to the new bounds in order to highlight the relativistic precessions with additional angular precession and eventually compare outcomes with GR's orbit.  
\newpage
\begin{figure}[!ht]
    \centering
    \subfigure[]{\includegraphics[width=0.44\textwidth]{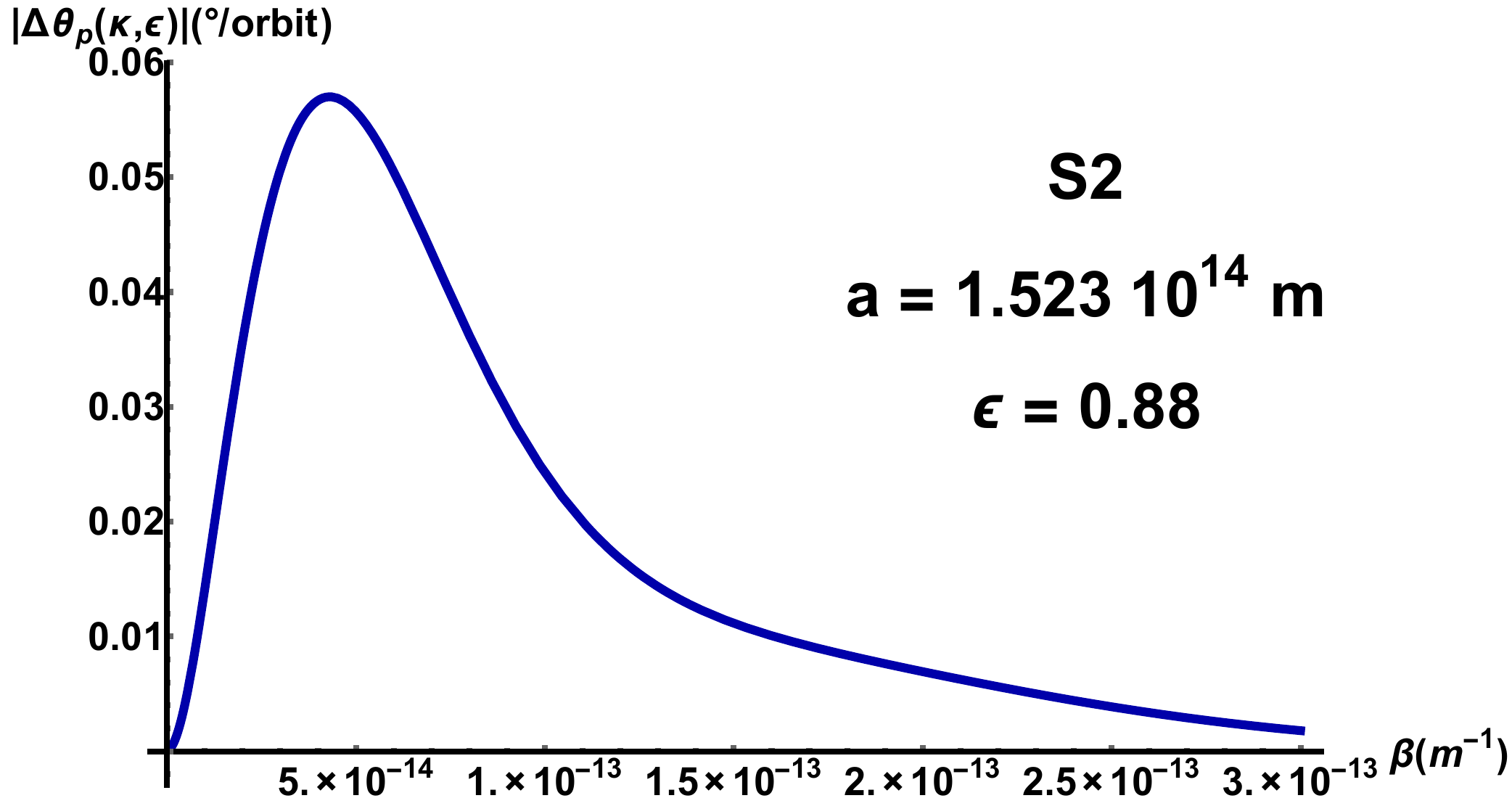}}
    \subfigure[]{\includegraphics[width=0.44\textwidth]{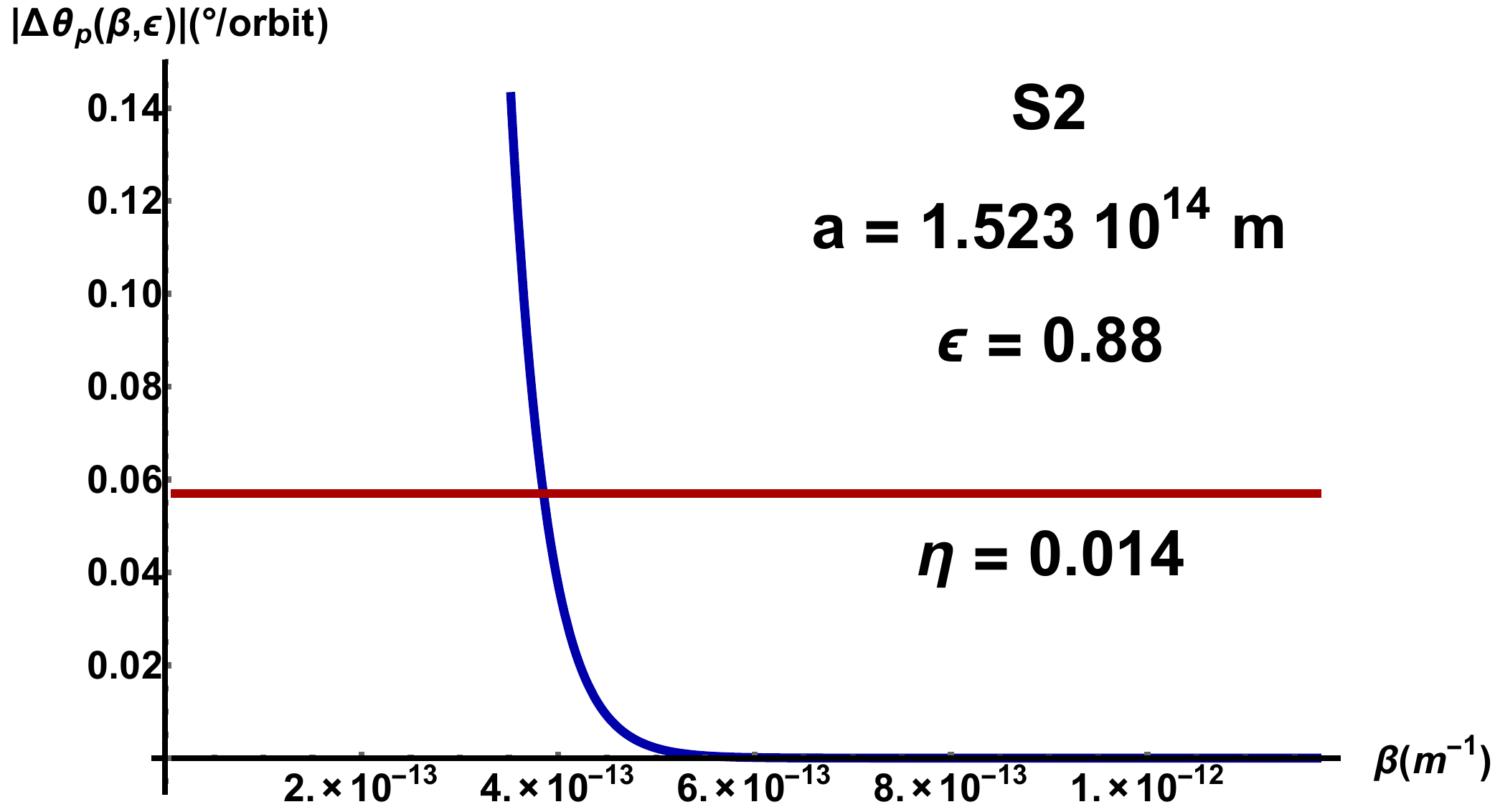}}
    \subfigure[]{\includegraphics[width=0.44\textwidth]{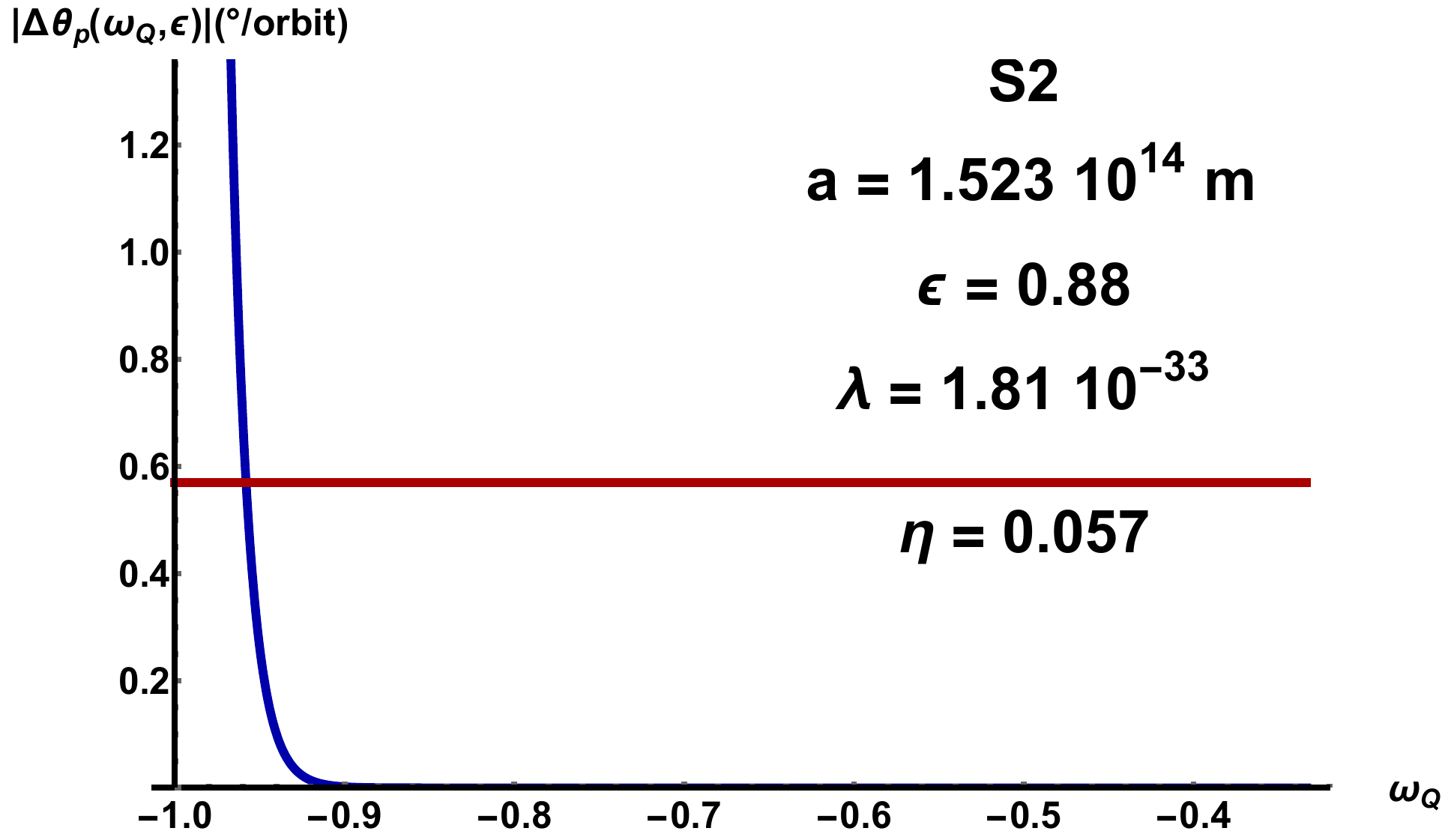}}
    \caption{(a) $|\Delta\theta_p(\kappa, \epsilon)|$ vs $\beta$ for S2 star in Scalar-Tensor-Fourth-Order Gravity (STFOG). (b) $|\Delta\theta_p(\beta, \epsilon)|$ vs $\beta$ for S2 star in NonCommutative Spectral Gravity (NCSG). (c) $|\Delta \theta_p(\omega_Q,\epsilon)|$ vs $\omega_Q$ for S2 star in Quintessence Field. $|\Delta \theta_p|$ precessions (blue line) and the constant $\eta$ (red line) plotted in $\degree/orbit$.}
    \label{NewfigS2}
\end{figure}
\newpage
\begin{itemize}
\item {\bf STFOG} - Referring to Scalar-Tensor Fourth Order Gravity, from Eq. (\ref{New-Fibound}) we obtain the new bound
  \begin{equation}\label{New-FiboundS2}
 |\Delta\theta_p(\kappa,\epsilon)|\lesssim \eta \quad \to \quad  |F_i|\, \sim  5.15\times 10^{-4} \,, \qquad i=\pm, Y\,.
 \end{equation}
In Fig. \ref{NewfigS2}(a), we have plotted the function $\Delta\theta_p(\kappa, \epsilon)$ for the S2 star. The maximum value of $\Delta\theta_p(\kappa, \epsilon)$ corresponding to $\beta \simeq 4.31 \times 10^{-14} \, \text{m}^{-1}$ (see Fig. \ref{NewfigS2}(a)) has been considered, while in Fig. \ref{orbsimSTFOG}, we illustrate the orbital simulation of the S2 star with respect to these values. The orbit exhibits a prograde rosette motion analogous to that of GR in Fig. \ref{orbsimGR} with a close value of the angular periastron precession.
\begin{figure}[!ht]
    \centering
    {\includegraphics[width=0.7665\textwidth]{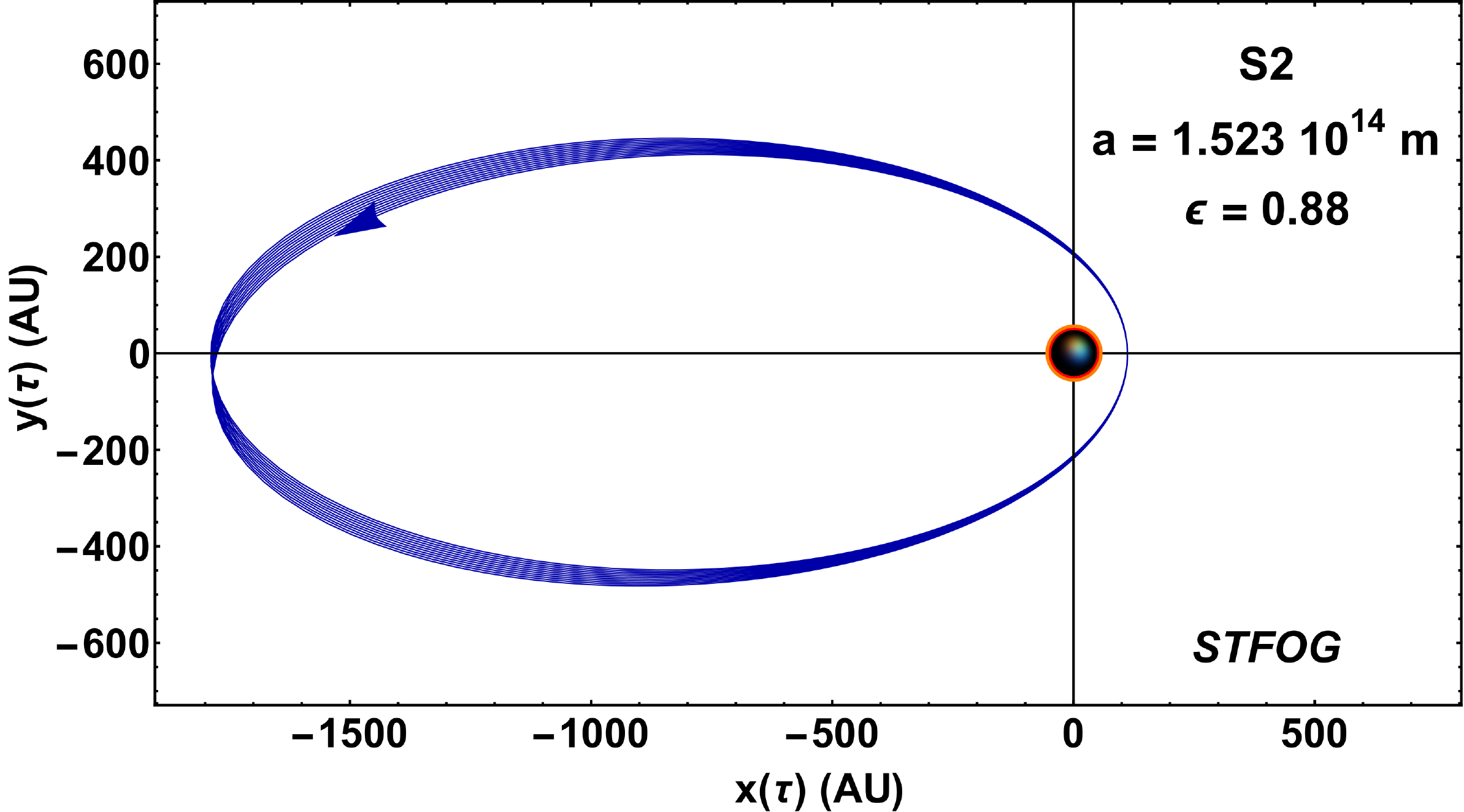}}
    \caption{Orbital simulation of the S2 star around Sgr A* predicted by 
 Scalar-Tensor-Fourth-Order Gravity (STFOG). The numerical solution of Eq. [\ref{Binet-equation}] is performed with respect to strength and range of interaction values $\beta \gtrsim 4.31 \times 10^{-14} \, \text{m}^{-1}$ and $|F_i| \sim 5.15\times 10^{-4}$ respectively. The axes are measured in astronomical units (AU). Sgr A* is oversized for better display.}
    \label{orbsimSTFOG}
\end{figure}

\item {\bf NCSG} - The S2 star values $\{\epsilon, \eta, a\}$, from (\ref{NewbetaNCSGbound}), imply that
 \begin{equation}\label{NewbetaNCSGboundS2}
 |\Delta \theta_p(\beta,\epsilon)|\lesssim \eta \quad \to \quad  |\beta|\lesssim \,\widetilde{\Theta}(\eta,\epsilon) \,.
 \end{equation}
 The result is reported in Fig. \ref{NewfigS2}(b). The further improved lower bound for $\beta$, with respect to $|\eta|=0.014$, is $\beta\gtrsim 5.43 \times 10^{-13}\, \text{m}^{-1}$, compatible with the astrophysical bounds in \cite{Nelson:2010rt, Nelson:2010ru} (Eq. [\ref{NCSG-periastron-advance}]).  In Fig. \ref{orbsimNCSG}, the orbital simulation of the S2 star for such a value is reported. Its prograde rosette motion highlights a periastron advance similar to the GR (Fig. \ref{orbsimGR}) as well, but a bit smaller because of the negative sign of the angular corrections to the total precession (see Eq. [\ref{NCSG-periastron-advance}]).

 \begin{figure}[!ht]
    \centering
   {\includegraphics[width=0.7665\textwidth]{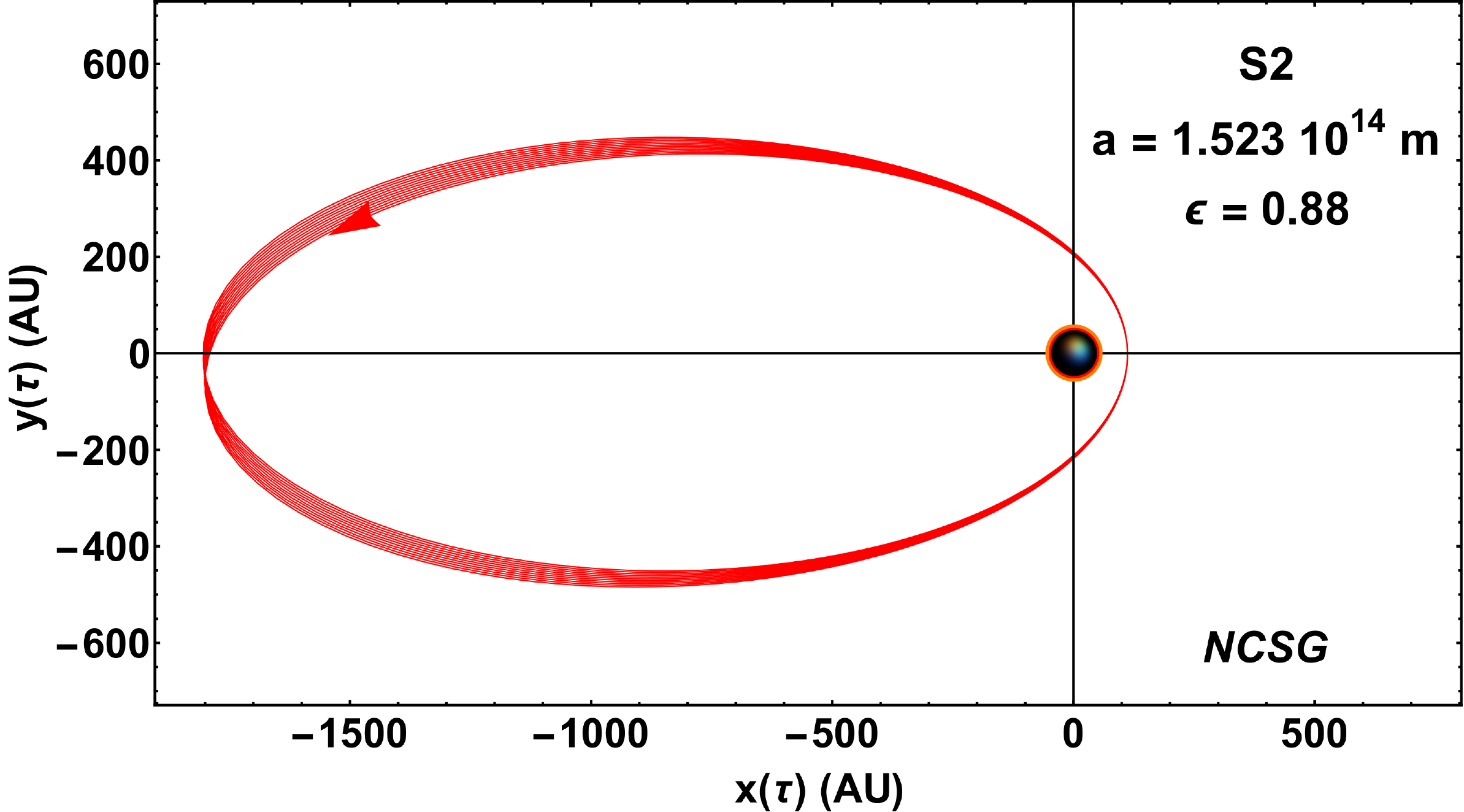}}
    \caption{Orbital simulation of the S2 star around Sgr A* predicted by 
 NonCommutative Spectral Geometry (NCSG). Numerical solution of the Eq. [\ref{Binet-equation}] performed by assuming the lower bound  to be $\beta\gtrsim 5.43 \times 10^{-13}\, \text{m}^{-1}$. Axes are measured in astronomical units (AU). Sgr A* is oversized for better display.}
    \label{orbsimNCSG}
\end{figure}
\newpage
 \item {\bf Quintessence} - In the case of Quintessence field deforming the Schwarzschild geometry, Eq. (\ref{precesspowerlaw2}) implies
  \begin{eqnarray}\label{precesspowerlawS2}
|\Delta \theta_p(\omega_Q, \epsilon)| \lesssim \eta \, .
\end{eqnarray}
We reported the results in Fig. \ref{NewfigS2}(c), from which it follows that for Quintessence $|\Delta \theta_p(\omega_Q, \epsilon)|\lesssim 0.057$ provided $\omega_Q \gtrsim -0.93$. Thus, the exact value $\omega_Q=-1$ that corresponds to the cosmological constant is excluded; the orbital simulation of S2 is finally reported in Fig. \ref{orbsimQF} with a behaviour close to that of the GR orbit (Fig. \ref{orbsimGR}). 

\begin{figure}[!ht]
    \centering
    {\includegraphics[width=0.7665\textwidth]{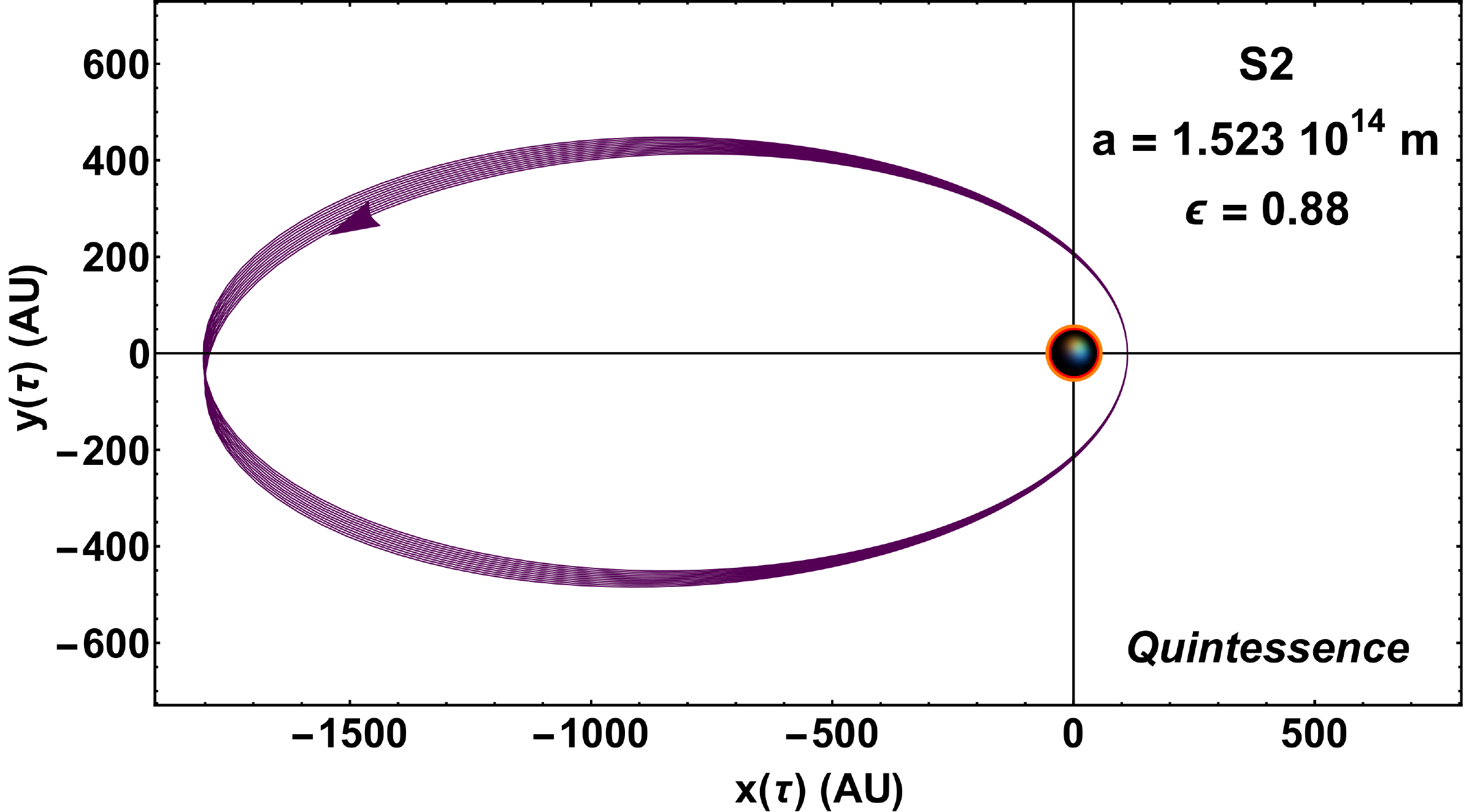}}    \caption{Simulation of the S2 star orbit around Sgr A* predicted by a Quintessence Field surrounding a Black Hole. The numerical solution of Eq. [\ref{Binet-equation}] is performed with respect to the bound $\omega_Q \gtrsim -0.93$ and $\lambda = 1.81 \times 10^{-33}$ (quintessential parameter) estimated from Eq. [\ref{precesspowerlawS2}]. The axes are measured in astronomical units (AU). Sgr A* is oversized for better display.}
    \label{orbsimQF}
\end{figure}
\end{itemize}

These new results generally lead to improvements in the constraints for curvature-based ETG, NCSG, and Quintessence Field models computed in our previous paper \cite{CapolupoLambiaseTedesco}. In particular, we notice that the sizes of the S2 orbits arising from numerical simulations are comparable with the orbit predicted by General Relativity and astronomical data. The relativistic periastron advance occurs in a prograde rosette motion for each theory, and the angular precessions are close to the general-relativistic value. Furthermore, it should be noted that the effects due to screening mechanisms, underlying the ETG models and operating on Earth and Solar System scales, could exist and be effective on larger scales, such as the galactic and extragalactic scales \cite{71,72,73,74}. Further observations over larger distances could provide limits on both screening mechanisms and higher derivative corrections, in particular, on the effective gravitational models discussed here. 
In this regard, new measurements on the S2 star orbit precession by the GRAVITY interferometer would be important in order to improve the level of accuracy, infer tighter constraints, and estimate the precise distance scale at which deviations from general-relativistic predictions become detectable. 

\section{Conclusions and remarks}
In this paper, after an epistemological introduction to the importance of anomalistic precessions in binary systems for the comprehension of new physics, we have studied the relativistic periastron advance beyond Einstein theory, in particular curvature-based Extended Theories of Gravity (ETG) and Quintessence Field have been considered. The gravitational interactions between massive bodies are thus described by extended/modified theories of gravity. In these models, the corrections to the Newtonian gravitational interaction are of the Yukawa-like form $V(r)=V_N(1+\alpha e^{-\beta r})$ (where $V_N=-GM/r$ is the Newtonian potential), or of the power-law form $V(r)=V_N+\alpha_q \, r^q$ in the case of Quintessential fields. The $2$-body system constitutes a good model for many astrophysical scenarios, such as those at the scale of Solar System, constituted by the Sun and a planet, as well as binary system composed by a Super Massive Black Hole and an orbiting star, which are both the most suitable candidates to test a gravitational theory. In particular, for Solar System planets and the S2 star around Sagittarius A*, we have dealt with a restricted version of the problem, namely systems which can be modelled as a test particle orbiting in central force field, i.e. moving along the geodesics of a Schwarzschild-like spherically symmetric space-time around a massive non-rotating ball-like source.

To this aim, we have found a new analytical solution through the epyciclic method, formally represented by a formula leading to a straightforward determination of the relativistic periastron advance beyond Einstein theory (Eq. [\ref{periastron-advance-in-ETG2}]) or models starting from the General Relativity's framework (e.g. GR plus dark matter). This includes the presence of all post-Newtonian potentials implied by the theories and it is also valid for large eccentricities, when deviations from the circular orbit are considerable. Eq. [\ref{additional-advance-in-ETG2}] is related to the total angular advance [\ref{periastron-advance-in-ETG2}], and provides the additional relativistic precession due to the post-Newtonian terms of the corrective potentials. Moreover, this solution reduces to its restricted version, given by Eqs. (\ref{periastron-advance-in-ETG})-(\ref{additional-advance-in-ETG}), when we consider systems with small orbital eccentricities; such a version turns out to be good for the Solar System. The generalisation of the result was based on the epicyclic method and the expansion to higher-order perturbations in the equation of orbit, which also involves relativistic corrections to the Newtonian potential. At the end of the process, indeed, we achieve an effective analytical formula that makes it immediately possible to calculate the orbital precession and includes all the post-Newtonian potentials beyond Einstein theory useful for analysing the dynamics of the system. By simply starting from the generic assumption of a spherical symmetric metric, such a resolution is universally valid and can be applied to analyse 2-body systems beyond General Relativity, or models within GR itself. Furthermore, it enables simple direct computations for the total precession, which is useful for high-precision gravitational tests. The results are analytical without the need for numerical integration, as might happen in other approaches. Afterwards, the main result (\ref{periastron-advance-in-ETG2})-(\ref{additional-advance-in-ETG2}) was directly applied to the Solar System and the S2 star in order to perform high-precision tests. The analysis for Scalar-Tensor-Fourth-Order Gravity (Eq. [\ref{STFOG-periastron-advance}]), NonCommutative Spectral Geometry (Eq. [\ref{NCSG-periastron-advance}]), and Quintessence Field related to dark energy (Eq. [\ref{precesspowerlaw2}]), has been performed to find improvements and, therefore, new constraints on the strength and range of interaction of such theories, because terms of relativistic origin can affect the final result. 

Thus, in the Solar System, we have found improvements leading to new bounds as follows: the highest value of $\beta_i$ is $\beta_i \simeq 3.54\times 10^{-11} \, \text{m}^{-1}$, with a constraint in $|F_i|$ being $|F_i| \gtrsim 3.44 \times 10^{-12}$.  In the case of Non-Commutative Spectral Gravity (NCSG), the analysis shows that the precession shift of planets allows us to constrain the parameter $\beta$ at $\beta > (10^{-11} - 10^{-10}) \, \text{m}^{-1}$. For the Quintessence Field, the adiabatic index $\omega_Q$ and the quintessence parameter $\kappa$ are the parameters that characterise the gravitational field; we have found that $\kappa$ assumes tiny values, as expected, being essentially related to the cosmological constant, while $\omega_Q\gtrsim -(0.78 - 0.69)$, that is, it never assumes the value $\omega_Q = -1$ corresponding to the pure cosmological constant.

For the S2 star around Sagittarius A*, we have found that for STFOG $\beta \gtrsim 4.31\times 10^{-14} \, \text{m} ^{-1}$, for NCSG we obtain $\beta \gtrsim 5.43 \times 10^{-13} \, \text{m}^{-1}$ compatible with astrophysical constraints, and finally for the Quintessence Field, we have $\omega_Q \gtrsim - 0.93$. Orbital simulations, numerically performed by assuming these bounds for the examined models, show a typical prograde rosette motion with relativistic periastron advance close to the GR's value. The sizes of the orbit predicted for STFOG, NCSG and Quintessence are comparable with the observed one traced on astrometric data and the one of Einstein theory itself. For NCSG, the dominant sign of the corrections is negative; then the prograde rosette motion performed by S2 has a tiny lower angle of precession than the general-relativistic one. For these theories, the S2 orbit simulations turn out to overall be consistent with the prograde rosette motion of the Einstein theory.
New constraints and simulation tests leading to further improvements will eventually be obtained when a new tighter error $\eta$ on the S2 star precession from the interferometer GRAVITY is available.

\section*{Acknowledgments}
The authors acknowledge the support of {\it Istituto Nazionale di Fisica Nucleare} (INFN).

\end{document}